\documentclass[aps,prd,superscriptaddress,twocolumn,nofootinbib]{revtex4-1}
\usepackage{graphicx}
\usepackage[caption=false]{subfig}
\usepackage{epstopdf}
\usepackage{amsmath}
\usepackage{amsfonts}
\usepackage{amssymb}
\usepackage{latexsym}
\usepackage{hyperref}
\usepackage[english]{babel}
\usepackage[utf8]{inputenc}
\usepackage[colorinlistoftodos]{todonotes}
\usepackage{color}
\usepackage{slashed}
\usepackage{feynmp}
\usepackage{bm}
\usepackage{bbold}
\usepackage{eufrak}
\usepackage{slashed}
\usepackage{bm}  
\usepackage{tabu}
\usepackage{xcolor}
\usepackage{tikz-feynman}
\usepackage{slashed}
\usepackage{ulem}
\usepackage{cancel}  

\begin{document}

\title{Phenomenology of a Born-Infeld extension of the $U(1)_{\rm Y}$ sector at lepton colliders}


\author{P. De Fabritiis}\email{pdf321@cbpf.br}
\affiliation{Centro Brasileiro de Pesquisas F\'{i}sicas (CBPF), Rua Dr Xavier Sigaud 150, Urca, Rio de Janeiro, Brazil, 22290-180}
	
\author{P. C. Malta}\email{pedrocmalta@gmail.com}
\affiliation{R. Antonio Vieira 23, 22010-100, Rio de Janeiro, Brazil}
	
\author{J. A. Helay\"el-Neto}\email{helayel@cbpf.br}
\affiliation{Centro Brasileiro de Pesquisas F\'{i}sicas (CBPF), Rua Dr Xavier Sigaud 150, Urca, Rio de Janeiro, Brazil, 22290-180}


\begin{abstract}
In this work we perform a non-linear extension of the $U(1)_{\rm Y}$ sector of the Standard Model leading to novel quartic effective interactions between the neutral gauge bosons. We study the induced effects through high-energy processes resulting in three photons, namely, Z-boson decay and electron-positron annihilation. Available experimental data on these processes do not yield viable lower bounds on the mass parameter $\sqrt{\beta}$, but we estimate that the range $\sqrt{\beta} \lesssim m_Z$ could be reliably excluded with better statistics in future $e^- e^+$ colliders. We also discuss neutral gauge-boson scatterings, contextualizing our findings with recent results on anomalous quartic gauge couplings.
\end{abstract}


\maketitle


\section{Introduction} \label{sec_intro}
\indent

The idea of non-linear electromagnetic responses of the vacuum was first suggested by Halpern~\cite{Halpern} and one year later by Heisenberg~\cite{Heisenberg}, where he proposed that virtual electron-positron pairs could be at the origin of photon-photon collisions. Soon thereafter, actions with non-linear electrodynamics were introduced by Born and Infeld~\cite{BI} and also Euler and Heisenberg~\cite{EulerHeisenberg} in the 1930s to deal with the classical problem of the infinite self-energy of a point charge. These extensions, which have been also explored in areas as diverse as black-hole physics and cosmology~\cite{BH1, BH2, cosmology1, cosmology2}, can display interesting features, such as vacuum birefringence and dichroism~\cite{Helayel1, Helayel2, Helayel3}. For recent developments, see Refs.~\cite{NonLinear1, NonLinear2, NonLinear3} and references therein.

Perhaps the most striking prediction of these models is the occurrence of light-by-light scattering already at tree level. This extremely rare process was recently observed by the ATLAS and CMS collaborations in heavy-ion collisions at the LHC~\cite{Light1, CMSlight, Light2}. The perspective to test effects of non-linear extensions of the Standard Model (SM) in high-energy experiments -- in lepton and hadron accelerators or potentially in photon colliders -- motivates us to search for possible phenomenological consequences.

The non-linear extension of traditional Maxwell electrodynamics modifies photon-photon interactions by introducing higher-order terms in the Lagrangian. Here, we are interested in extending the whole hypercharge sector of the electroweak gauge group, thus giving rise to other interesting phenomena. In fact, besides reproducing the already known non-linear effects in standard electrodynamics (corrected by a factor involving the Weinberg angle), this extension induces anomalous quartic couplings between the Z-boson and the photon. This in turn theoretically allows for rare processes to take place already at tree level, as for example the creation of a Z-boson pair from the collision of two photons.


There is also a more recent motivation for considering non-linear models. Introduced by Dirac 90 years ago~\cite{Dirac}, magnetic monopoles remain elusive despite experimental efforts. It was thought for a long time that it would be impossible to obtain a monopole solution in the electroweak sector because of its gauge structure after symmetry breaking, but this belief turned out to be wrong. An electroweak monopole solution was obtained by Cho and Maison~\cite{CM}, but the original solution predicted an infinite mass that should be regularized to have physical meaning and sustain any hope of being found experimentally. A few years ago, some proposals of SM extensions regularizing the monopole energy and giving a finite, calculable mass were made~\cite{CKY, Ellis1, Regcharge}. A non-linear extension of the hypercharge sector could solve the infinite-energy problem~\cite{Arunasalam}; a more general extension of the $U(1)_{\rm Y}$ sector giving a finite-energy monopole solution was investigated in Ref.~\cite{Philipe}. Nowadays there is hope to finally find a monopole in dedicated experiments, such as MoEDAL at CERN~\cite{MOEDAL2}, so it is imperative to understand the phenomenological implications of such an extension.

Non-linear effects have not been observed at low energies. This means that the parameter controlling the non-linearity of the fields is expected to be large in comparison to other relevant energy scales. The parameters in non-linear theories may be constrained in different ways, {\it e.g.}, via hydrogen spectroscopy or interferometry~\cite{beta1, beta2}.  A more stringent bound is obtained using LHC data on light-by-light scattering in heavy-ion collisions~\cite{Ellis2}. The lower bound reported there is $\sim 100 \, \text{GeV}$, but it could reach $\sim 200 \, \text{GeV}$ under less restrictive assumptions. The
ATLAS data on $gg \rightarrow \gamma \gamma$ can enhance this sensitivity by	1 order of magnitude in a Born-Infeld (BI) extension of SM~\cite{Ellisgluon}, reaching	the TeV scale as in brane-inspired models.

In this work, we analyze an extension of the hypercharge sector of the SM. This gives rise to quartic effective interactions between the neutral gauge bosons absent in the SM at tree level. These novel operators contribute to decay and scattering processes and we explore existing experimental data to place lower bounds on the non-linear parameter. We discuss recent results constraining anomalous gauge couplings and briefly consider possible improvements on these bounds in future experiments.

This paper is organized as follows: in Sec.~\ref{sec_model}, we present the theoretical setup of our model. In Sec.~\ref{sec_limits}, we discuss options to constrain the expansion parameter $\beta$, in particular through the decay $Z \rightarrow 3 \, \gamma$ in Sec.~\ref{sec_Z3gamma} and the scattering $e^- \, e^+ \rightarrow 3 \, \gamma$ in Sec.~\ref{sec_eeAAA}. In Sec.~\ref{sec_gauge_scats}, we discuss neutral gauge-boson scattering processes and contextualize our discussion with recent results in anomalous quartic gauge couplings. Finally, in Sec.~\ref{sec_conclusion}, we present our closing remarks. We use natural units ($c = \hbar = 1$) and the flat Minkowski metric $\eta^{\mu\nu}={\rm diag}(+1,-1,-1,-1)$ throughout.

\section{Theoretical setup} \label{sec_model}
\indent

Let us briefly review the usual electroweak (EW) Lagrangian to fix our notation. The bosonic part of the EW sector is given by
\begin{align}\label{key}
\mathcal{L}_{\text{EW}} &= \mathcal{L}_{\text{Gauge}} + \mathcal{L}_{\text{Higgs}},
\end{align}
where
\begin{align}
\mathcal{L}_{\text{Gauge}} &= -\frac{1}{4} F_{\mu \nu}^a F_{\mu \nu}^a -\frac{1}{4} B_{\mu \nu} B_{\mu \nu},  \\
\mathcal{L}_{\text{Higgs}} &= \vert D_\mu H \vert^2 - \lambda \left( H^\dagger H - \frac{m^2}{2 \lambda}\right)^2 \, .
\end{align}
Here we defined the covariant derivative as
\begin{equation} \label{cov_der}
D_\mu = \partial_\mu - i g A_\mu^a T^a - i g' Y B_\mu.
\end{equation}
In the equations above, $A_\mu^a$ and $B_\mu$ are the gauge fields associated with the gauge group $SU(2)_{\rm L} \times U(1)_{\rm Y}$, $F_{\mu \nu}^a = \partial_\mu A_\nu^a - \partial_\nu A_\mu^a + g \epsilon_{abc} A_\mu^bA_\nu^c$ and $B_{\mu \nu} = \partial_\mu B_\nu - \partial_\nu B_\mu$ are the respective field strengths, and $g$ and $g'$ are the couplings. Here, $T^a$ are the generators of $SU(2)_{\rm L}$ satisfying $\left[T^a , T^b\right] = i \epsilon^{abc} T^c$, and $Y$ is the weak hypercharge.

The Higgs field $H$ is a $SU(2)_{\rm L}$ doublet with hypercharge $Y(H) = +1/2$. The scalar potential induces a non-trivial vacuum expectation value given by $\vert \langle H \rangle \vert^2 = v^2/2 = m^2/2\lambda$. Below this energy scale, the theory is cast into the Higgs phase with three massive vector bosons $W^\pm$, $Z$, a massive scalar $h$ and a massless photon A ($\gamma$) in the spectrum. The physical fields can be written using the Weinberg angle $\theta_W$: the neutral vector bosons are defined by $Z_\mu = \cos\theta_W \, A_\mu^3 - \sin\theta_W \, B_\mu$ and $A_\mu = \sin\theta_W \, A_\mu^3 + \cos\theta_W \, B_\mu$, whereas the charged vector fields are defined by $W^{\pm}_\mu = \left( A_\mu^1 \mp i A_\mu^2 \right)/\sqrt{2}$.

The masses of the vector bosons can be precisely measured and are $m_W = gv/2 = 80.4$~GeV and $m_Z = m_W/\cos\theta_W =  91.2$~GeV. The Weinberg angle can be experimentally determined and satisfies $\sin^2\theta_W = 0.23$. After symmetry breaking, the kinetic part of the gauge Lagrangian (omitting mass terms) reads
\begin{align}\label{gaugekin}
\mathcal{L}_{\rm Gauge}^{\rm Kin} = - \frac{1}{4} \, F_{\mu \nu} F^{\mu \nu} - \frac{1}{4} \, Z_{\mu \nu} Z^{\mu \nu}
- \frac{1}{2} \, W_{\mu \nu }^+ W^{\mu \nu -} \, ,
\end{align}
where the field-strength tensors are defined as usual.


We may now introduce the leptons through the following Lagrangian:
\begin{align}\label{lag_lep}
\mathcal{L}_{\text{Leptons}} = i \, \bar{L}_i \gamma^{\mu} D_\mu L_i + i \, \bar{\ell}_{iR} \gamma^{\mu} D_\mu \ell_{iR} \, ,
\end{align}
where $L_i$ denotes the lepton doublets $L_i = \left( \nu_{iL} \,\,\, \ell_{iL}  \right)^t$ with $\nu_{iL}$, $\ell_{iL}$, and $\ell_{iR}$ representing the left-handed neutrinos, the left-handed charged leptons, and the right-handed lepton fields, respectively. Here, $i$ is a flavor index to distinguish between the three generations of leptons. The hypercharge assignment adopted here is $Y(L_i) = -1/2 $ and $Y(\ell_{iR}) = -1$. Taking Eq.~\eqref{lag_lep} with Eq.~\eqref{cov_der}, including the gauge fields after symmetry breaking, we obtain the interactions between matter and gauge fields. In what follows, only two such interaction terms will be relevant, namely,
\begin{eqnarray}
\mathcal{L}_{ee\gamma} &=& - e  \bar{\ell}_{i} \gamma_{\mu} \ell_{i}  A^{\mu} \label{lag_eeA} \, , \\ 
\mathcal{L}_{eeZ} &=& \frac{g}{4 \cos\theta_W}  \bar{\ell}_{i} \gamma_{\mu} \left( -1 + 4\sin^2\theta_W + \gamma^5  \right) \ell_{i}  Z^{\mu} \label{lag_eeZ} \, .
\end{eqnarray}


Here we propose the following general extension of the weak hypercharge sector of the EW Lagrangian,
\begin{align}\label{key}
\mathcal{L} = - \frac{1}{4} B_{\mu \nu} B^{\mu \nu} \quad \longrightarrow \quad  \mathcal{L}_{\rm Y} =  f \left( \mathcal{F}, \mathcal{G}  \right) \, ,
\end{align}
where we defined the Lorentz and gauge invariant objects 
\begin{equation}
\mathcal{F} = \frac{1}{4} B_{\mu \nu} B^{\mu \nu} \quad \quad {\rm and} \quad \quad \mathcal{G} = \frac{1}{4} B_{\mu \nu} \tilde{B}^{\mu \nu}
\end{equation}
with the dual field-strength tensor given by $\tilde{B}^{\mu \nu} = \frac{1}{2} \epsilon^{\mu \nu \rho \sigma} B_{\rho \sigma}$. This type of non-linear extension was already studied in the context of magnetic monopoles~\cite{Philipe}, where it was shown that under certain conditions, it allows a finite-energy electroweak monopole solution. 

The SM predictions are so far in excellent agreement with experiment, and in order to recover the usual SM results, we demand that our general extension $f \left( \mathcal{F}, \mathcal{G}  \right)$ reproduces the usual term $- \frac{1}{4} B_{\mu \nu} B^{\mu \nu}$ in some appropriate limit. Since we do not want to have a parity-violating term in the photon sector after spontaneous symmetry breaking, we impose the physically motivated assumption that $f \left( \mathcal{F}, \mathcal{G}  \right)$ depends on $\mathcal{G}$ only through $\mathcal{G}^2$.

Let us consider for instance a BI extension of the hypercharge sector~\cite{BI} given by
\begin{align}\label{lag_BI}
\mathcal{L}_{\rm Y}^{\text{BI}} = \beta^2\left[ 1 - \sqrt{1 + 2 \left( \frac{\mathcal{F}}{\beta^2} -\frac{\mathcal{G}^2}{2 \beta^4}\right)  }\right] \, ,
\end{align}
where $\beta$ is a parameter with dimension of mass squared. This non-linear extension has been extensively studied in the context of electrodynamics, with applications in a range of subjects, and has attracted a lot of interest in the recent years after the observation of light-by-light scattering at the LHC~\cite{Light1, CMSlight, Light2}. Interestingly enough, the BI action can be derived from string theory~\cite{BIstring} and also appears in the dynamics of D-branes~\cite{BIbrane}.

Our goal is to study the phenomenological consequences of the non-linear extension in high-energy processes. To accomplish this, we need to obtain the induced operators written in terms of the physical fields after symmetry breaking. The mass scale set by $\sqrt{\beta}$ is expected to be large in comparison with the typical energies of the processes considered, motivating us to perform a Taylor expansion of Eq.~\eqref{lag_BI} in powers of $X = \frac{\mathcal{F}}{\beta^2} -\frac{\mathcal{G}^2}{2 \beta^4}$:
\begin{equation}\label{nonlinearlag}
\mathcal{L}_{\rm Y} = -\mathcal{F} +  \frac{1}{2 \beta^2} \left[ \mathcal{F}^2 + \mathcal{G}^2 \right]  + \mathcal{O}\left(1/\beta^4\right) \, .
\end{equation}
We will only consider tree-level processes with at most four gauge bosons in each vertex, so we can safely restrict ourselves to leading non-trivial order. It is important to keep in mind that this perturbative approach can only be trusted as long as the energy of the process is lower than the mass scale set by $\sqrt{\beta}$, as this guarantees that the next terms in the expansion provide increasingly negligible corrections to the leading-order terms.


Furthermore, taking into consideration the recent interest in different versions of non-linear electrodynamics, we can also consider other interesting extensions that would give rise to the same physical effects in the approximation considered here. In fact, using $X$ defined above, we could as well have considered here the extensions given by  $\mathcal{L}_{\rm Y} ^{\text{Log}} = - \beta^2 \log\left[1 + X \right] $ and $\mathcal{L}_{\rm Y} ^{\text{Exp}} = \beta^2 \left[ e^{-X } -1 \right]$  giving us the $U(1)_{\rm Y}$ version of the logarithmic~\cite{Helayel1} and exponential~\cite{Helayel2, Helayel3} non-linear electrodynamics. The three extensions agree up to leading non-trivial order and we will restrict ourselves to tree-level processes with at most four gauge bosons interactions, so we may safely consider the $\beta$ parameters as being equal with a good approximation and use Eq.~\eqref{nonlinearlag} to describe their common effects.

The Lagrangian above is a function of the $U(1)_{\rm Y}$ gauge field, $B_\mu$, but after symmetry breaking, we can write it in terms of the physical fields, $A_\mu$ and $Z_\mu$, retrieving the usual SM kinetic terms at zeroth order. At first order, we have ($s_\theta \equiv \sin\theta_W, c_\theta \equiv \cos\theta_W$)
\begin{eqnarray}\label{lag_ZA}
	\mathcal{L}_{\rm Y}^{(1/\beta^2)} & = &  \frac{1}{32 \beta^2} \bigg\{ s_\theta^4 \left[ (ZZ)(ZZ) + (Z\tilde{Z})(Z\tilde{Z}) \right]  
	\nonumber \\
	& + &  c_\theta^4 \left[ (FF)(FF) + (F\tilde{F})(F\tilde{F}) \right]  \nonumber \\
	& + & 2 s_\theta^2 c_\theta^2  \left[ (FF)(ZZ) + (F\tilde{F})(Z\tilde{Z}) \right]  \nonumber \\
	& + & 4 s_\theta^2 c_\theta^2  \left[ (ZF)(ZF) + (Z\tilde{F})(Z\tilde{F}) \right]  \nonumber \\
	& - & 4  s_\theta^3 c_\theta  \left[ (ZZ)(ZF) + (Z\tilde{Z})(Z\tilde{F}) \right] \nonumber \\
	& - & 4  s_\theta c_\theta^3  \left[ (FF)(FZ) + (F\tilde{F})(F\tilde{Z}) \right] \bigg\} \, ,
\end{eqnarray}
where we defined $\left(ZZ\right) \equiv Z_{\mu \nu} Z^{\mu \nu}$ with an analogous definition for the dual versions. All non-linearly induced vertices above have the same momentum structure and very similar Feynman rules; this traces back to the common origin of such interactions.

In conclusion, we see that our non-linear extension in the hypercharge sector generates a series of dimension-8 effective operators generically suppressed by a factor $\left(\mathcal{E}/\Lambda\right)^4$, where $\mathcal{E}$ is a typical energy scale characteristic of the process and $\Lambda$ is the mass scale set by $\sqrt{\beta}$. These effective operators will introduce new vertices, allowing processes that could only occur in the SM at loop level to take place already at tree level. In the next section we explore this fact and consider different high-energy processes to obtain lower bounds on $\beta$ whenever experimental data are available. We also discuss the impact of our non-linear extension on scattering processes involving neutral gauge bosons.

	
\section{Experimental limits} \label{sec_limits}
\indent

In the section above we have extracted quartic interaction vertices between the photon and Z-boson which are completely absent from the SM, thus opening up interesting possibilities to constrain the expansion parameter $\beta$. In the following we explore a few of them.

	
\subsection{$Z \rightarrow 3\, \gamma$} \label{sec_Z3gamma}
\indent

In the SM there is no tree-level $Z\gamma\gamma\gamma$ vertex, so the decay process $Z \rightarrow 3\, \gamma$ proceeds exclusively via fermion and W-boson loops~\cite{boudjema, pham}. The theoretical prediction for the partial width is $\Gamma\left( Z \rightarrow 3\, \gamma \right)_{\rm SM}  = 1.4$~eV~\cite{Z3gamma_th} , and given the experimentally determined total width of the Z-boson $\Gamma_{\rm exp}^{\rm Z} = 2.49$~GeV~\cite{PDG}, the expected branching ratio is ${\rm BR}\left( Z \rightarrow 3\, \gamma \right)_{\rm SM} = 5.4 \times 10^{-10}$. The currently best upper bound was obtained by the ATLAS Collaboration using pp collisions at $\sqrt{s} = 8$~TeV and reads~\cite{ATLAS_3gamma}
\begin{equation} \label{BR_Z3gamma}
{\rm BR}\left( Z \rightarrow 3\, \gamma \right)_{\rm exp} < 2.2 \times 10^{-6} \, ,
\end{equation}
representing a five-fold improvement on the previous determination from LEP~\cite{L3}. This process is clearly very rare and could not yet be measured directly. It is thus an ideal testing ground for new physics~\cite{Villa, Velasco}.

\begin{figure}[t!]
\begin{minipage}[b]{.5\linewidth}
\includegraphics[width=\textwidth]{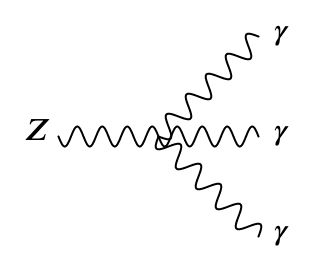}
\end{minipage} \hfill
\caption{Tree-level Feynman diagram for the decay $Z \rightarrow 3\, \gamma$.}
\label{fig_ZAAA}
\end{figure}

The SM prediction is compatible with the best current experimental bound, but there is a vast gap between them. The non-linear extension can therefore be constrained by comparing its prediction to the experimental bound; cf. Eq.~\eqref{BR_Z3gamma}. The tree-level amplitude for a Z-boson with 4-momentum $p$ decaying into three photons with 4-momenta $q_k$ ($k$ = 1,2,3) is (cf. Fig.~\ref{fig_ZAAA})
\begin{equation} \label{amp_Z3gamma}
-i \mathcal{M} = \epsilon_\alpha (p) V^{\alpha\beta\gamma\delta}_{Z3\gamma}\left( \beta \right) \epsilon_\beta^* (q_1) \epsilon_\gamma^* (q_2) \epsilon_\delta^* (q_3) \, ,
\end{equation}
where the vertex factor 
\begin{equation} \label{vert_factor_ZAAA}
V^{\alpha\beta\gamma\delta}_{Z3\gamma} \left( \beta \right) = -i \frac{s_\theta c_\theta^3}{\beta^2} f^{\alpha\beta\gamma\delta}
\end{equation}
may be read from the last line of Eq.~\eqref{lag_ZA}. The momentum-dependent function $f^{\alpha\beta\gamma\delta}$ is given by
\begin{eqnarray} \label{eq_f}
-f^{\alpha\beta\gamma\delta} & = & \left[ (q_1 \cdot q_2) \eta^{\beta\gamma} - q_1^\gamma q_2^\beta  \right] \left[ (p \cdot q_3) \eta^{\alpha\delta} - p^\delta q_3^\alpha  \right]      \nonumber \\
& + & \left[ (q_1 \cdot q_3) \eta^{\beta\delta} - q_1^\delta q_3^\beta  \right] \left[ (p \cdot q_2) \eta^{\alpha\gamma} - p^\gamma q_2^\alpha  \right]      \nonumber \\
& + & \left[ (q_2 \cdot q_3) \eta^{\gamma\delta} - q_2^\delta q_3^\gamma  \right] \left[ (p \cdot q_1) \eta^{\alpha\beta} - p^\beta q_1^\alpha  \right]      \nonumber \\
& + & \epsilon^{\mu\beta\rho\gamma} \epsilon^{\nu\delta\kappa\alpha} p_{\kappa} q_{1 \mu} q_{2 \rho} q_{3 \nu}   \nonumber \\
& + & \epsilon^{\mu\beta\rho\delta} \epsilon^{\nu\gamma\kappa\alpha} p_{\kappa} q_{1 \mu} q_{2 \nu} q_{3 \rho}   \nonumber \\
& + & \epsilon^{\mu\gamma\rho\delta} \epsilon^{\nu\beta\kappa\alpha} p_{\kappa} q_{1 \nu} q_{2 \mu} q_{3 \rho}  \, .
\end{eqnarray}
Here we have assumed that $p$ flows into the vertex, whereas the $q_k$ flow out of it. Incidentally, this momentum structure is the same for all vertices in Eq.~\eqref{lag_ZA}.

From this point on, we neglect the loop-level SM amplitude so the tree-level result from Eq.~\eqref{amp_Z3gamma} is essentially the only contribution to the decay. The unpolarized squared amplitude reads
\begin{equation} \label{amp2_Z3gamma}
\langle |\mathcal{M}|^2 \rangle = \frac{8 s^2_\theta c_\theta^6}{3 \beta^4} \Phi(p,q_1,q_2,q_3) \, ,
\end{equation}
with the momentum factor given by
\begin{eqnarray} \label{PHI_0}
\Phi(p,q_1,q_2,q_3) & = & \frac{1}{2} \left(p \cdot q_1\right)^2 \left(q_2 \cdot q_3\right)^2  + {\rm perm.}  \, ,
\end{eqnarray}
where ``perm." indicates all permutations of the $q_k$. In the rest frame of the decaying Z-boson, $p^\mu = (m_Z, 0)$, and the outgoing photons have $E_k = \vert {\bf q}_k \vert$. By applying the usual dispersion relations and momentum conservation, we find 
\begin{equation}\label{key}
p \cdot q_3 = m_Z E_3 \quad {\rm and} \quad q_1 \cdot q_2 = \frac{m^2_Z}{2} - m_Z E_3 \, ,
\end{equation}
with similar results for the other 4-momenta pairs. Therefore, we can rewrite $\Phi(p,q_1,q_2,q_3)$ as
\begin{eqnarray}\label{PHI}
\Phi(p,q_1,q_2,q_3)  =  \frac{m_Z^4}{4} \sum\limits_{k=1,2,3}  E_k^2  (m_Z - 2 E_k)^2 \, .
\end{eqnarray}

Notice that this expression is symmetric under the change of final photons, a reasonable behavior since there is no preferred photon in this decay. As the phase space integral also enjoys this symmetry, we can simply use one of the terms above to do the integration and multiply the output by 3, since they will necessarily give the same result. The partial width is defined as
\begin{equation}
d\Gamma = \frac{1}{3!} \frac{1}{2m_Z} \langle \vert \mathcal{M} \vert^2 \rangle  d\Pi_3 \, ,
\end{equation}
where $1/3!$ is the the symmetry factor due to the identical photons in the final state. The three-body phase-space function is 
\begin{eqnarray}\label{PI_3}
d\Pi_3 & = & \frac{d^3 {\bf q}_1}{(2 \pi)^3 \, 2 E_1} \frac{d^3 {\bf q}_2}{(2 \pi)^3 \, 2 E_2} \frac{d^3 {\bf q}_3}{(2 \pi)^3 \, 2 E_3} \nonumber \\
& \times & (2 \pi)^4 \delta^4\left(p - q_1 - q_2 - q_3\right) \, .
\end{eqnarray}
We have then
\begin{eqnarray}\label{key}
d\Gamma & = & K \frac{E_3 (m_Z - 2E_3)^2}{E_1 E_2} d^3 {\bf q}_1 d^3 {\bf q}_2 d^3 {\bf q}_3   \nonumber \\
& \times & \delta^4\left(p - q_1 - q_2 - q_3\right), 
\end{eqnarray}
where the constant K, already including the factor of 3, is
\begin{align}\label{key}
K = \frac{s_\theta^2 c_\theta^6 m_Z^3}{1536 \pi^5 \beta^4} \, .
\end{align}

The rest of the calculation follows a path similar to the textbook calculation of muon decay~\cite{Stohr}. The delta function may be split into two factors enforcing energy and 3-momentum conservation. The latter allows us to write ${\bf q}_2 \rightarrow -\left( {\bf q}_1 + {\bf q}_3 \right)$ and $E_2 \rightarrow \vert {\bf q}_1 + {\bf q}_3 \vert$. Let us take the polar axis along ${\bf q}_3$, which is held fixed, so that 
\begin{eqnarray}
E_2(\cos\theta) & = & \vert {\bf q}_1 + {\bf q}_3 \vert  \nonumber \\
& = & \sqrt{ E_1^2 + E_3^2 + 2E_1 E_3 \cos\theta } \, .
\end{eqnarray}
We may then write $d^3 {\bf q}_1 = 2\pi E_1^2 d\vert {\bf q}_1 \vert d\cos\theta$, and we have
\begin{eqnarray} \label{key}
d\Gamma & = & 2\pi K \frac{E_1 E_3 (m_Z - 2 E_3)^2}{ \vert {\bf q}_1 + {\bf q}_3 \vert} d^3 {\bf q}_3 dE_1 d\cos\theta \nonumber \\  
& \times & \delta\left[ g(\cos\theta) \right] \, ,
\end{eqnarray}
where $g(\cos\theta) = m_Z - E_1 - E_2(\cos\theta) - E_3$.

Now, the delta function cannot be directly integrated, so we need to change variables. This redefinition leads to
\begin{equation}
\delta\left[ g(\cos\theta) \right] = \frac{ E_2(\cos\theta) }{E_1 E_3} \delta\left( \cos\theta - \cos\theta_0 \right) \, ,
\end{equation}
where $\cos\theta_0$ is such that $g(\cos\theta_0) = 0$. The delta function now implies that both the maximum energy of any individual photon and the minimum energy of any pair of photons are $m_Z/2$. Consequently, we have $E_1$ and $E_3$ limited to the ranges $(\frac{m_Z}{2} - E_3, \, \frac{m_Z}{2} )$ and $(0, \, \frac{m_Z}{2})$, respectively. Performing the remaining integrations and dividing by the Z-boson width we find that the branching ratio is given by
\begin{eqnarray}\label{Gamma_Y}
{\rm BR} \left( Z \rightarrow 3\, \gamma \right)_{\rm Y} & = & \frac{s_\theta^2 c_\theta^6 }{184320 \,\pi^3 \, \Gamma_{\rm exp}^{\rm Z}}  \frac{m_Z^9}{\beta^4} \nonumber \\
& = & 6.7 \times 10^{-7} \, \left( \frac{m_Z}{\sqrt{\beta}} \right)^8 \, .
\end{eqnarray}

We are finally able to place an experimental bound on $\beta$. The branching ratio predicted by the SM is extremely small ($\sim 10^{-10}$), way below current experimental sensitivities. Allowing the result above to fully saturate the experimental upper limit, {\it i.e.}, ${\rm BR} \left( Z \rightarrow 3\, \gamma \right)_{\rm Y} \simeq {\rm BR} \left( Z \rightarrow 3\, \gamma \right)_{\rm exp}$; cf. Eq.~\eqref{BR_Z3gamma}. This implies that
\begin{equation} \label{lim_ZAAA}
\sqrt{\beta} \gtrsim 80 \, {\rm GeV} \, ,
\end{equation}
which is slightly lower than the bound reported in Ref.~\cite{Ellis2}. In Ref.~\cite{helayel_mario}, the authors adopt the result of Eq.~\eqref{lim_ZAAA} above on the BI parameter to make estimates on the redshift and to discuss birefringence and dichroism in connection with a class of p-extended BI-type actions in the presence of an external uniform magnetic field.

Here we must add an important remark. The energy scale of a decay process is set by the mass of the decaying particle, here given by $m_Z = 91.2$~GeV. Therefore, the bound obtained above must be taken with a grain of salt since it represents a mass scale lower than the energy of the process, challenging the basic assumption behind our effective-theory approach. Nonetheless, it is worth noticing that this restriction is a matter of experimental limitation: the best bound on the Z-decay into three photons is still orders of magnitude away from the SM prediction, so we may confidently expect that future experiments will yield much more stringent bounds on it, therefore significantly improving on the result above.

\begin{figure}[t!]
\begin{minipage}[b]{1.\linewidth}
\includegraphics[width=\textwidth]{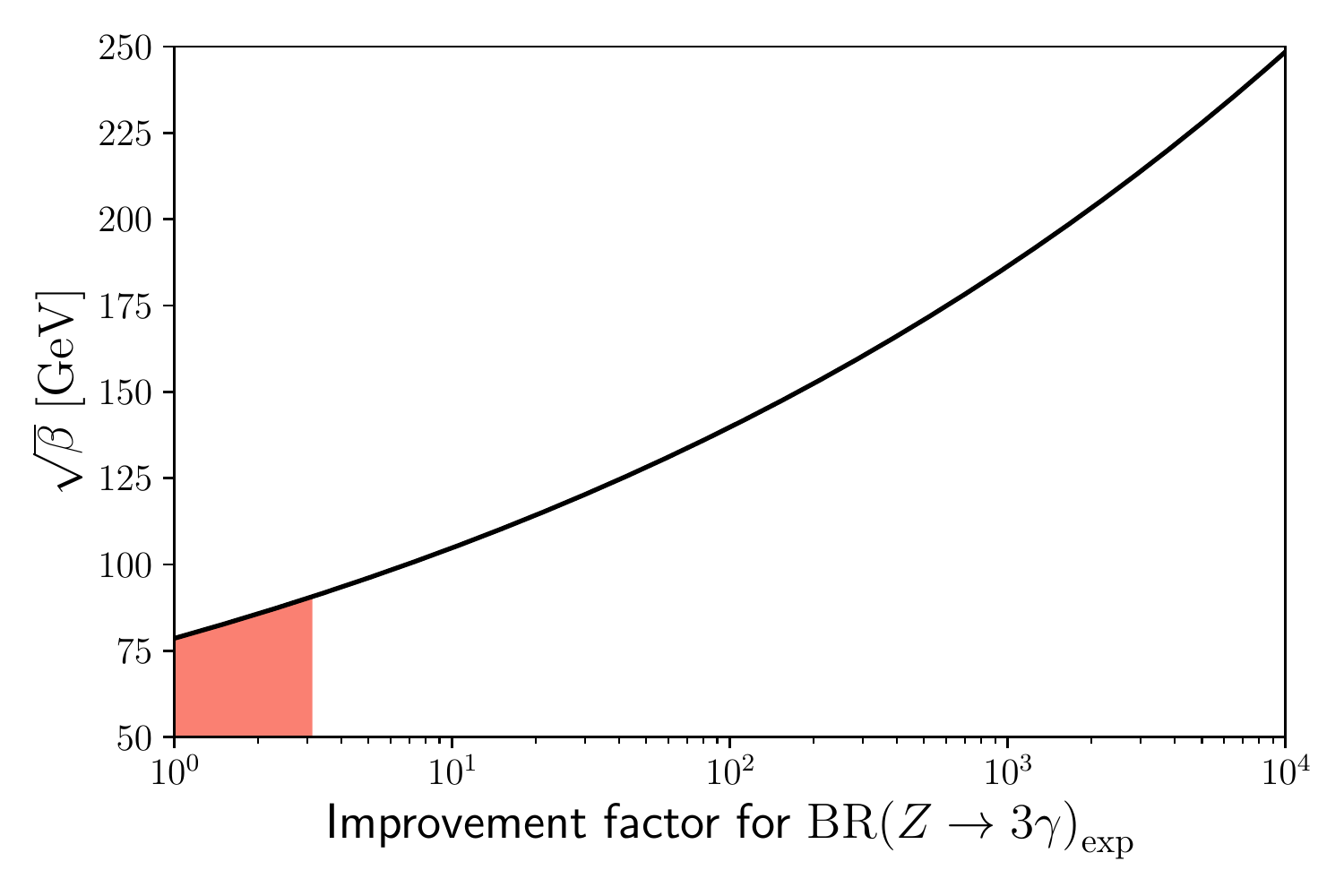}
\end{minipage} \hfill
\caption{Projection for the lower bound on $\sqrt{\beta}$ as a function of the improvement factor of the experimental sensitivity relative to the currently best one; cf. Eq.~\eqref{BR_Z3gamma}~\cite{ATLAS_3gamma}. Incidentally, Eq.~\eqref{Gamma_Y} reaches the order of magnitude of the SM prediction with $\sqrt{\beta} \sim 220$~GeV. The region shaded in red is such that $\sqrt{\beta} < m_Z$, where our predictions are not accurate.}
\label{fig_beta_ZAAA}
\end{figure}

The lower bound in Eq.~\eqref{lim_ZAAA} is clearly limited by the experimental sensitivity. If the current experimental upper bound on the branching ratio [cf.~Eq.~\eqref{BR_Z3gamma}] would be improved by a factor of $\sim 3$ -- a smaller improvement than the one from ATLAS~\cite{ATLAS_3gamma} relative to LEP~\cite{L3} -- we would be able to exclude the region $\sqrt{\beta} \lesssim m_Z$. Future lepton colliders, {\it e.g.} , ILC~\cite{ilc, white, Fujii_1, Fujii_2} and FCC-ee~\cite{fcc1, fcc2}, whose main goal is precision Higgs physics, could operate at the Z-pole and produce a vast sample of Z-bosons; the ILC and the FCC-ee could produce, respectively, $10^2$ and $10^5$ times more Z-bosons than LEP. It is therefore possible, with much better statistics and improved detector capabilities, to improve the upper limit on ${\rm BR} \left( Z \rightarrow 3\gamma \right)$ enough to constrain $\sqrt{\beta}$ at or above $m_Z$.

In Fig.~\ref{fig_beta_ZAAA}, we plot the lower bound on $\sqrt{\beta}$ as a function of the future improvement of the experimental sensitivity, ${\rm BR}\left( Z \rightarrow 3\, \gamma \right)_{\rm exp}$, relative to the currently best one~\cite{ATLAS_3gamma}. The situation discussed in the paragraph above is illustrated by the area shaded in red; an improvement of at least $\sim 3$ would lead to viable bounds. The unfortunately weak dependence of the expansion parameter on the experimental sensitivity is made explicit by the slope of the curve, meaning that only large improvements in sensitivity would lead to noticeable improvements in the lower bound on our non-linear extension.

As a final remark we note that the discussion above relies on the fact that, so far (and in the foreseeable future), only upper limits on the process $Z \rightarrow 3\, \gamma$ could be placed. The SM prediction is 4 orders of magnitude below the currently best upper bound, so we may also speculate about possible limits on the expansion parameter in case the SM expectation is eventually confirmed. In this scenario, there is no tension between the SM and experiment, so we may assume that the non-standard result is responsible for a small correction of the SM prediction, being hidden under the (relative) experimental uncertainty, {\it i.e.}, ${\rm BR}_{\rm Y}(\beta)/{\rm BR}_{\rm SM} \lesssim \delta_{\rm exp}$. Conservatively assuming $\delta_{\rm exp} \sim 10\%$ would allow us to improve the lower bound to $\sqrt{\beta} \gtrsim 295$~GeV. For even better precisions of $1\%$ and $0.1\%$ we find $\sqrt{\beta} \gtrsim 395$~GeV and $\sqrt{\beta} \gtrsim 530$~GeV, respectively.

	
\subsection{$e^- \, e^+ \rightarrow 3 \, \gamma$} \label{sec_eeAAA}
\indent

Hadron colliders have played a central role in the establishment of the SM as our best theory of elementary particles and their interactions; great examples are the discoveries of the W- and Z-bosons, as well as of the Higgs scalar. However, lepton colliders, such as LEP, were crucial in subsequent precision measurements, helping to probe not only tree-level predictions, but also radiative corrections~\cite{wim}. The next development is to achieve even higher precision in measurements of electroweak parameters, in particular those related to the Higgs and gauge bosons~\cite{tadeusz}.

Lepton colliders represent optimal tools to this end, and next-generation machines have been proposed, such as ILC~\cite{ilc, white, Fujii_1, Fujii_2}, FCC-ee~\cite{fcc1,fcc2}, CEPC~\cite{cepc}, and CLIC~\cite{clic1}. These are designed to study the SM in great detail, but searching for deviations from the SM that could hint at new physics is an equally important goal. In this context, the process $e^- \, e^+ \rightarrow 3 \, \gamma$ offers an interesting option to test modifications of the gauge couplings, in particular those involving photons and Z-bosons. From Eq.~\eqref{lag_ZA}, we see that our non-linear extension induces precisely such anomalous couplings that could give rise to new contributions for processes with three photons in the final state. We note that the SM contribution is very well described by QED with negligible electroweak corrections.

\begin{figure}[t!]
\begin{minipage}[b]{1.\linewidth}
\includegraphics[width=\textwidth]{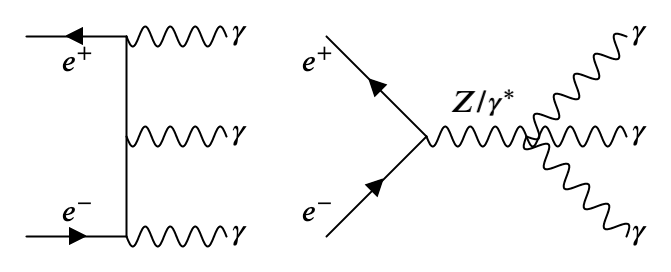}
\end{minipage} \hfill
\caption{The lowest-order Feynman diagrams contributing to the scattering $e^- (p_1) \, e^+ (p_2) \rightarrow \gamma (q_1) \, \gamma (q_2) \, \gamma (q_3)$.}
\label{fig_eeAAA}
\end{figure}

The Feynman diagrams contributing to $e^- \, e^+ \rightarrow 3 \, \gamma$ at tree level are shown in Fig.~\ref{fig_eeAAA}. The QED contribution is given by
\begin{eqnarray}
& - & i \mathcal{M}_{\rm QED} = ie^3 \overline{v}(p_2) \Bigg[ \gamma^\rho \frac{\slashed{p}_1 - \slashed{q}_1 - \slashed{q}_2}{\left( p_1 - q_1 - q_2 \right)^2}  \gamma^\nu   \nonumber \\
 &\times  &  \frac{\slashed{p}_1 - \slashed{q}_1}{\left( p_1 - q_1 \right)^2} \gamma^\mu \Bigg] u(p_1) \epsilon_\mu^* (q_1) \epsilon_\nu^* (q_2) \epsilon_\rho^* (q_3)  \, , \label{amp_eeAAA_0} 
\end{eqnarray}
which must be added to the other five amplitudes obtained from this one by permutation of the external photons. We are considering high-energy scatterings, so the electron mass may be safely neglected. The non-linearly induced photon- and Z-mediated amplitudes are given by
\begin{eqnarray} 
-i \mathcal{M}_\gamma & = & \frac{-e}{\left( p_1 + p_2 \right)^2} \overline{v}(p_2) \gamma_\mu u(p_1)  \nonumber \\
& \times & V^{\mu\nu\beta\rho}_{4\gamma}\left( \beta \right)  \epsilon_\nu^* (q_1) \epsilon_\beta^* (q_2) \epsilon_\rho^* (q_3) \, , \label{amp_eeAAA_1} \\
-i \mathcal{M}_{\rm Z} & = & \frac{g_Z}{\left( p_1 + p_2 \right)^2 - m_Z^2 + im_Z \Gamma_Z} \nonumber \\
& \times & \overline{v}(p_2) \gamma_\mu \left( c_v - c_a \gamma^5  \right) u(p_1) \nonumber \\
& \times & V^{\mu\nu\beta\rho}_{Z3\gamma}\left( \beta \right) \epsilon_\nu^* (q_1) \epsilon_\beta^* (q_2) \epsilon_\rho^* (q_3) \, , \label{amp_eeAAA_2}
\end{eqnarray}
where $g_Z = e/4s_\theta c_\theta$, $c_v = -1 + 4s_\theta^2$ and $c_a = -1$. The Z-width is $\Gamma_Z = 2.49$~GeV. The $Z\gamma\gamma\gamma$ vertex was defined in Eq.~\eqref{vert_factor_ZAAA}, and the four-photon vertex is analogous:
\begin{equation} \label{vert_factor}
V^{\alpha\beta\gamma\delta}_{4\gamma} \left( \beta \right) = i \frac{c_\theta^4}{\beta^2} f^{\alpha\beta\gamma\delta} \, .
\end{equation}
The function $f^{\alpha\beta\gamma\delta}$ is given by Eq.~\eqref{eq_f} with the appropriate relabeling of the 4-momenta.

The total tree-level amplitude for the process, $\mathcal{M}$, is $\mathcal{M} = \mathcal{M}_{\rm QED} + \mathcal{M}_{\gamma} + \mathcal{M}_{\rm Z}$ and the total unpolarized cross section is given by
\begin{equation} \label{cs_eeAAA_0}
d \sigma = \frac{1}{3!} \frac{1}{2 E_{\rm cm}^2} \langle \vert \mathcal{M} \vert^2 \rangle d\Pi_3 \, ,
\end{equation}
where $1/3!$ is the the symmetry factor due to the identical photons in the final state and the phase-space factor is the same as in Eq.~\eqref{PI_3}. The squared amplitude is essentially the sum of three contributions: a pure QED part, an interference term between QED and the non-linear amplitudes, and a purely non-linear term. The contribution from pure QED is discussed in Appendix~\ref{app_A}.

Let us first discuss the interference term, $\langle \vert \mathcal{M} \vert^2_{\rm QED-Y} \rangle $. The unpolarized squared amplitude is quoted in detail in Appendix~\ref{app_B}. To simplify matters, we may express all energies and 3-momenta in units of the c.m. energy, $E_{\rm cm}$, so that we can write it as 
\begin{equation} \label{eq_amp_interf_0}
\langle \vert \mathcal{M} \vert^2_{\rm QED-Y} \rangle = \mathcal{X}_{\rm QED-Y} e^4 c_\theta^2 \frac{E_{\rm cm}^2}{\beta^2} \mathcal{A}(p_i, q_j) \, ,
\end{equation}
with $\mathcal{A}(p_i, q_j)$ representing a function of the now dimensionless energies and 3-momenta that the reader may obtain from Eq.~\eqref{eq_app_interf}. The pre-factor $\mathcal{X}_{\rm QED-Y}$ is given by ($x=m_Z^2 / E_{\rm cm}^2$ and $y = \Gamma_Z^2 / m_z^2$)
\begin{equation}\label{eq_X}
\mathcal{X}_{\rm QED-Y}  = \frac{ 3 - (3 + 4c_\theta^2)x + 4c_\theta^2 x^2 (1 + y) }{ (1 - x)^2 + yx^2 } \, .
\end{equation}


From the phase-space volume, we get another factor of $E_{\rm cm}^2$ that cancels the one present in the denominator of Eq.~\eqref{cs_eeAAA_0}, so that, putting all the numerical factors together, we finally obtain
\begin{equation} \label{cs_eeAAA_interf}
\sigma\left( e^- e^+ \rightarrow 3 \gamma \right)_{\rm QED-Y} = \mathcal{X}_{\rm QED-Y}  \frac{\alpha^2 c_\theta^2}{384 \pi^3} \frac{s}{\beta^2} \, \mathcal{I}_{\rm QED-Y} 
\end{equation}
with $s = E_{\rm cm}^2$, $e^2 = 4\pi\alpha$ and 
\begin{equation} \label{eq_I}
\mathcal{I}_{\rm QED-Y} = \int \mathcal{A}(p_i, q_j) \frac{d^3 {\bf q}_1}{E_1} \frac{d^3 {\bf q}_2}{E_2} \frac{d^3 {\bf q}_3}{E_3} \delta^4\left(\Sigma p_i - \Sigma q_j \right).
\end{equation}
Note that the quantities in Eq.~\eqref{eq_I} are all expressed in units of $\sqrt{s} = E_{\rm cm}$, being therefore dimensionless.

Equation~\eqref{eq_I} cannot be easily evaluated analytically due to the complexity of the integrand, so we solve it numerically via standard Monte Carlo methods. The Dirac delta enforcing 4-momentum conservation severely constrains the phase-space volume available to the outgoing photons. In fact, their individual energies are bound to be at most $0.5$ and the sum of any pair of energies must be larger than this value, allowing us to limit the range of the sampled 3-momentum components to the interval $\left[ -0.5, 0.5 \right]$. In what follows, we use data from $e^- e^+$ collisions at LEP resulting in two or three photons and the cross sections quoted were obtained under the experimental conditions of the detector. That means that we have to impose similar cuts to our theoretical cross sections if we want to compare them to LEP data.

Particularly important are the angular and energy cuts imposed~\cite{L3_1992, L3_2002}. Since the forward-backward direction along the beam is inaccessible, the range in polar angles is limited to $16^\circ \leq \theta_\gamma \leq 164^\circ$, {\it i.e.}, the detectable photons must satisfy $|\cos\theta_\gamma| \leq 0.96$ to be contained in the electromagnetic calorimeter. Furthermore, the individual photons must have an energy $E_\gamma > 5$~GeV. Even though Eq.~\eqref{eq_I} is written in terms of dimensionless quantities, the aforementioned lower threshold on the detectable energy of the single photons introduces an energy dependence, as the cut is expressed as $E_\gamma > 5/\sqrt{s}$. The values of the integral evaluated at selected energy values are quoted in Table~\ref{table_integrals}. For the sake of concreteness, the interference cross section at $\sqrt{s} = 207$~GeV is
\begin{equation} \label{cs_eeAAA_interf_207}
\sigma_{\rm QED-Y} \left( \sqrt{s} = 207 \, {\rm GeV} \right) \simeq 0.88 \left( \frac{250 \, {\rm GeV}}{\sqrt{\beta}}  \right)^4 {\rm fb} \, .
\end{equation}


We now move on to the the purely non-linear contribution, $\langle |\mathcal{M}|^2_{\rm Y} \rangle$, which is expected to be sub-dominant relative to the interference term discussed above. The unpolarized squared amplitude is stated in Eq.~\eqref{eq_app_Y}, and after expressing the 4-momenta in units of $E_{\rm cm}$, we have
\begin{equation} \label{amp_eeAAA_Y}
\langle \vert \mathcal{M} \vert^2_{\rm Y} \rangle = \mathcal{X}_{\rm Y} e^2 c_\theta^4 \frac{E_{\rm cm}^6}{\beta^4} \mathcal{B}(p_i, q_j) \, ,
\end{equation}
with $\mathcal{B}(p_i, q_j)$ representing a dimensionless function in analogy with $\mathcal{A}(p_i, q_j)$. The pre-factor is
\begin{equation} \label{KCcoeff}
\mathcal{X}_{\rm Y} =  \frac{5  - 12c_\theta^2 x + 8 c_\theta^4 x^2(1 + y)}{(1 - x)^2 + yx^2} \, .
\end{equation}


Equation~\eqref{amp_eeAAA_Y} may be integrated analytically\footnote{The result without detector cuts is $\sigma_{\rm Y} = \mathcal{X}_{\rm Y}  \frac{\alpha c_\theta^4}{368640 \pi^2} \frac{s^3}{\beta^4}$.}, but here we adopted the same Monte Carlo setup employed in the calculation of the interference term. The cross section is then given by
\begin{equation} \label{cs_eeAAA_Y}
\sigma \left( e^- e^+ \rightarrow 3 \gamma \right)_{\rm Y} =  \mathcal{X}_{\rm Y} \frac{\alpha c_\theta^4}{6144 \pi^4} \frac{s^3}{\beta^4} \, \mathcal{I}_{\rm Y} \, ,
\end{equation}
where $\mathcal{I}_{\rm Y}$ is defined analogously to $\mathcal{I}_{\rm QED-Y}$; cf. Eq.~\eqref{eq_I}. Specializing to $\sqrt{s} = 207$~GeV and using the numerical value of the integral including detector cuts from Table~\ref{table_integrals}, we have
\begin{equation} \label{cs_eeAAA_Y207}
\sigma_{\rm Y} \left( \sqrt{s} = 207 \, {\rm GeV} \right) \simeq  0.01 \left(  \frac{250 \, {\rm GeV}}{\sqrt{\beta}} \right)^8  {\rm fb} \, .
\end{equation}

In the discussion above we have obtained the total cross sections involving the novel neutral vertices originating in Eq.~\eqref{lag_ZA}. The fact that only quartic vertices are produced implies that $e^- \, e^+ \rightarrow 2 \, \gamma$ does not receive corrections, at least at tree level, but $e^- \, e^+ \rightarrow 3 \, \gamma$ does. From dimensional analysis alone, we expect the number of events with two photons to be roughly 2 orders of magnitude times larger than with three photons, thus making dedicated searches for three-photon events harder. Therefore, more commonly, experiments look for multi-photon processes and the best available data to our knowledge were collected at LEP where the c.m. energy of the $e^- \, e^+$ pair was scanned passing by the Z-pole and reaching more than 200~GeV.

\setlength{\tabcolsep}{4pt}
\renewcommand{\arraystretch}{0.9}
\newcolumntype{C}[1]{>{\centering\arraybackslash}m{#1}}
\begin{table}[]
\centering
\begin{tabular}[t]{@{}|C{1.3cm}||C{1.0cm}|C{1.0cm}|C{1.0cm}|C{1.0cm}|C{-0.6cm}@{}}
\cline{1-5}
$\sqrt{s}$~(GeV)  & 91.2 & 207 & 250 & 350 &      \\  [6pt]
\cline{1-5}
$\mathcal{I}_{\rm QED}$ & 27006 & 37976 & 41796 & 45854 &    \\ 
[13pt]
\cline{1-5}
$\mathcal{I}_{\rm QED-Y}$ & 19.45 & 20.02 & 20.13 & 20.24 &    \\ 
[13pt]
\cline{1-5}
$\mathcal{I}_{\rm Y}$ & 0.138 & 0.139 & 0.139 & 0.139  &      \\ 
[13pt]
\cline{1-5}
\end{tabular}
\captionsetup{justification=centering}
\caption{Values of the numerical integrals appearing in Eqs.~\eqref{cs_eeAAA_interf}, \eqref{cs_eeAAA_Y},  and~\eqref{I_qed}. The first two energy values are relevant in the context of existing LEP data~\cite{L3_1992, L3_2002}, whereas the last two are important for future $e^- e^+$ colliders, such as the ILC~\cite{ilc, white, Fujii_1, Fujii_2, lum1}. The following cuts were applied: $E_\gamma > 5$~GeV and $|\cos\theta_\gamma| < 0.96$~\cite{L3_1992, L3_2002}.}
\label{table_integrals}
\end{table}

The L3 Collaboration analyzed LEP data of events resulting in multiphoton final states~\cite{L3_1992, L3_2002}. Since electroweak corrections are heavily suppressed, these measurements provide a clean test of QED, whose predictions were successfully confirmed. The calculations of the QED expectation were performed following Ref.~\cite{Berends}, where contributions up to $\mathcal{O}(\alpha^3)$ are considered, {\it i.e.}, the tree-level cross sections for two and three final photons plus radiative corrections. Here, however, we are working with an effective theory and we limit our analysis to tree level and we refrain from employing their results.

The tree-level cross section for $e^- \, e^+ \rightarrow 2 \, \gamma$ is well known; cf. Eq.~\eqref{cs_qed_AA}. No closed form for the tree-level cross section for $e^- \, e^+ \rightarrow 3 \, \gamma$ in the CM could be found, so we calculated the squared amplitude analytically and performed the phase-space integration numerically including the appropriate detector cuts; cf. Eq~\eqref{cs_qed_AAA}. Let us consider concrete data to try to constrain $\sqrt{\beta}$. Since we are dealing with an effective theory whose effects grow with energy, we will ignore data at the Z-pole~\cite{L3_1992} and focus on the more promising high-energy results~\cite{L3_2002}.

The L3 collaboration analyzed $e^- \, e^+ \rightarrow \gamma \gamma (\gamma)$ data in detail and indicates cross-section measurements for final states with two and three photons. The highest energy bin is 207~GeV (cf. Table~3 of Ref.~\cite{L3_2002}), and it quotes the expected $\mathcal{O}(\alpha^3)$ cross section as 9.9~pb, whereas our tree-level result is~9.2~pb. Given that the difference includes radiative contributions deliberately unaccounted for here and possible effects from further selection criteria, we are confident that our calculation delivers a meaningful result for the QED prediction at tree-level.

Now, given that QED accurately describes the experimental data, we are only able to find lower bounds on $\sqrt{\beta}$. In fact, we may constrain it by demanding that the effects of the non-linear extension hide under the relative experimental uncertainty
\begin{equation}  \label{eq_constraint_eeAA}
\frac{ \sigma_{\rm QED-Y} + \sigma_{\rm Y}  }{ \sigma_{\rm QED} } \lesssim \delta_{\rm exp} \, ,
\end{equation}
with $\sigma_{\rm QED}$ being the tree-level expectation from QED. The cross sections for final states with two and three photons are, respectively, $\sigma^{2\gamma}_{\rm QED}$, Eq.~\eqref{cs_qed_AA}, and $\sigma^{3\gamma}_{\rm QED}$, Eq~\eqref{cs_qed_AAA}. For the sake of concreteness, we focus on the highest energy bin quoted in Table~3 from Ref.~\cite{L3_2002}, $\sqrt{s} = 207$~GeV, for which the relative uncertainty of the measured cross section is $\delta_{\rm exp} = 0.34/10.16 \simeq 0.034$. Plugging this and $\sigma^{2\gamma}_{\rm QED} + \sigma^{3\gamma}_{\rm QED}  = 9.2$~pb into Eq.~\eqref{eq_constraint_eeAA}, we obtain $\sqrt{\beta} \gtrsim 73$~GeV.

The absolute number of $e^- \, e^+ \rightarrow 3 \, \gamma$ events is also reported in Ref.~\cite{L3_2002} for different energies, albeit without the respective experimental uncertainties. Focusing again on $\sqrt{s} = 207$~GeV, the expected tree-level cross section for pure QED is $0.29$~pb. At this energy, 29 three-photon events were observed, so we may conservatively assume that the uncertainty is $\sim \sqrt{29} \simeq 5.4$ events. Taking into account the effective integrated luminosity, $87.8$~pb$^{-1}$, this is equivalent to $0.06$~pb, so that the relative uncertainty is $\delta_{\rm exp} = 0.06/0.29 \simeq 0.21$. With $\sigma_{\rm QED}  = \sigma^{3\gamma}_{\rm QED} = 0.29$~pb, Eq.~\eqref{eq_constraint_eeAA} gives $\sqrt{\beta} \gtrsim 97$~GeV.

The bounds found above suffer from the same limitation as the one from the analysis of Z-decay: $\sqrt{\beta} < \sqrt{s}$. This is, however, not surprising, since the experimental uncertainties are orders of magnitude larger than the typical values expected from Eqs.~\eqref{cs_eeAAA_interf_207} and~\eqref{cs_eeAAA_Y207}. We are thus confronted with the fact that the currently available data on $e^- \, e^+ \rightarrow \gamma \gamma (\gamma)$ do not yield viable bounds on $\sqrt{\beta}$.

Despite being experimentally more challenging, measuring $e^- \, e^+ \rightarrow 3 \, \gamma$ has the largest potential, as only the process directly affected by the non-linear effects is examined. We conclude, therefore, that a sensible lower limit on $\sqrt{\beta}$ could be placed if future $e^- \, e^+$ colliders would include measuring this process in their research programs. Let us take the ILC as an example, which targets a total integrated luminosity of $14$~ab$^{-1}$ over its full operation time~\cite{lum1}. For the sake of clarity, let us focus on the initial stage with $\sqrt{s} = 250$~GeV, where an integrated luminosity of $\sim 500$~fb$^{-1}$ is planned to be attained in the first five years. Assuming similar detector cuts as at LEP and a (pessimistic) $1\%$ effective luminosity\footnote{For comparison, the analysis of $e^- \, e^+ \rightarrow \gamma \gamma (\gamma)$ at LEP in the energy range 192 -- 209~GeV contained 0.43~fb$^{-1}$ of data, roughly ten times less.}, $\sim 5$~fb$^{-1}$, pure QED predicts 1073 three-photon events, whereas the non-linear terms would contribute with three extra events for $\sqrt{\beta} = 300$~GeV; {\it i.e.}, the level of precision required would be $3/1073 \sim 0.3\%$. A similar precision would be required at $\sqrt{s} = 350$~GeV with $\sqrt{\beta} = 400$~GeV.

\subsection{Pure gauge-boson scatterings} \label{sec_gauge_scats}
\indent

The electroweak sector of the SM is based on the non-Abelian gauge group $SU(2)_{\rm L} \times U(1)_{\rm Y}$. This is manifest in the form of the covariant derivative; cf.~Eq.~\eqref{cov_der}, and the non-linear transformation properties of the gauge bosons. Particularly relevant is the presence of triple and quartic self-interaction couplings in the gauge sector. As a matter of fact, in a pure Yang-Mills theory, the quartic coupling is related to the triple one, even at the quantum level, as a consequence of gauge symmetry -- this is a trademark feature of a non-Abelian gauge theory. Given that the structure of the gauge self-couplings in the electroweak sector is completely specified by construction, any deviations from this would suggest the presence of new physics.

Measurements of the gauge self-couplings are therefore especially interesting from both theoretical and experimental points of view. Particularly important are high-energy scattering processes involving the Z-boson and the photon, which could give a clear signal indicating SM extensions modifying the hypercharge sector like the one proposed here. With this in mind, we consider some of the possible scattering processes proceeding via the quartic couplings from Eq.~\eqref{lag_ZA} already at tree level, instead of loop-level as predicted by the SM.

As mentioned in Sec.~\ref{sec_eeAAA}, $e^- e^+$ colliders offer clean conditions for precision tests of the SM. More interestingly, there are currently proposals of machines that can be adapted to work as linear photon colliders. Important sources of photons at a linear lepton collider include bremsstrahlung~\cite{Yellin} and Compton laser back-scattering~\cite{Kim} (there is also beamstrahlung~\cite{Friberg}). At LEP or LHC bremsstrahlung is the dominating form of radiation production, whereas at TESLA~\cite{tesla}, ILC~\cite{ilc, ilc2} or CLIC~\cite{clic1, AA_CLIC1, AA_CLIC2}, Compton back-scattering of electrons in intense lasers would be used to produce $\gamma\gamma$ or $e\gamma$ collisions. 
In this scenario, the photons created may carry a substantial amount of the electron energy~\cite{Telnov}.

Given that future linear $e^- e^+$ machines envision in their prospects the possibility of an extension to include photon colliders at relatively low cost, let us focus on $\gamma \gamma$ collisions producing exclusively vector bosons $V_i = \gamma, \, Z, \, W^\pm$. In this context, measuring, {\it e.g.}, the process $\gamma \gamma \rightarrow W^{+} W^{-}$ in a photon collider is an attractive option due to its large ($\sim$80~pb) cross section~\cite{Denner95, Yehudai}. The non-linear realization of the hypercharge sector proposed in this work, however, does not affect charged gauge bosons, so we shall focus on $\gamma \gamma$ fusion leading to neutral gauge bosons as final products: $\gamma \, \gamma \rightarrow \gamma \, Z$, $\gamma \, \gamma \rightarrow Z \, Z$, and $\gamma \, \gamma \rightarrow \gamma \, \gamma$. It is noteworthy that, within the SM framework, these processes receive only loop-level contributions, but here they will be induced at tree level by the effective operators present in Eq.~\eqref{lag_ZA}.

For the sake of concreteness, in the following we compute the non-linear contribution to the unpolarized cross section of the process $\gamma \, \gamma \rightarrow \gamma \, Z$ at tree level. Though we consider this particular process in more detail, all others may be analyzed by similar means. From Eq.~\eqref{lag_ZA}, we see that the relevant vertex factor is $V^{\alpha\beta\gamma\delta}_{Z3\gamma}$, cf. Eqs.~\eqref{vert_factor_ZAAA} and~\eqref{eq_f}, but with the substitutions: $p \rightarrow p_1$, $q_1 \rightarrow -p_2$, $q_2 \rightarrow q_1$ and $q_3 \rightarrow q_2$ appropriate for a 2-to-2 scattering.

\begin{figure}[t!]
\begin{minipage}[b]{.5\linewidth}
\includegraphics[width=\textwidth]{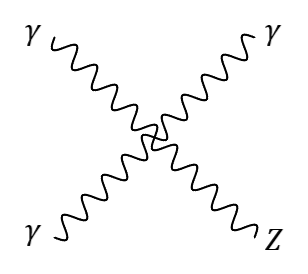}
\end{minipage} \hfill
\caption{The lowest-order Feynman diagram contributing to the scattering $\gamma (p_1) \, \gamma (p_2) \rightarrow \gamma (q_1) \, Z (q_2)$.}
\label{fig_AAAZ}
\end{figure}

The tree-level amplitude for this process is then
\begin{equation} \label{amp_AAAZ}
	-i \mathcal{M} = \epsilon_\alpha (p_1) \epsilon_\beta(p_2) V^{\alpha\beta\gamma\delta}_{Z3\gamma}\left( \beta \right)  \epsilon_\gamma^* (q_1) \epsilon_\delta^* (q_2) 
\end{equation}
with the momenta attributions given in Fig.~\ref{fig_AAAZ}. Here, we are assuming that the unpolarized photons are on shell and monochromatic\footnote{This is a simplified scenario, and a more detailed analysis would follow the strategy from Ref.~\cite{Ellis2}, for example.}. After summing and averaging over polarizations, the unpolarized squared amplitude becomes
\begin{align}
	\langle \vert \mathcal{M} \vert^2 \rangle &=	\frac{c_\theta^6 s_\theta^2}{8 \beta^4} \left[m_Z^4 \left(s^2+t^2+u^2\right) \right. \nonumber \\ 
	& \left. -2 m_Z^2 \left(s^3+t^3+u^3\right) + s^4+t^4+u^4\right],
\end{align}
%
where the Mandelstam variables, expressed in terms of the c.m. energy $E_{\rm cm}$ of the incoming photons and the scattering angle $\theta$, are
\begin{subequations}
	\begin{eqnarray}\label{mandelstam_AAAZ}
		s & = & E_{\rm cm}^2 \; , \\
		t & = & - \frac{1}{2} (E_{cm}^2 - m_Z^2) (1 - cos\theta) \; , \\
		u & = & -\frac{1}{2} (E_{cm}^2 - m_Z^2) (1 + cos\theta) \; .
	\end{eqnarray}
\end{subequations}
%

\begin{figure}[t!]
\begin{minipage}[b]{1.0\linewidth}
\includegraphics[width=\textwidth]{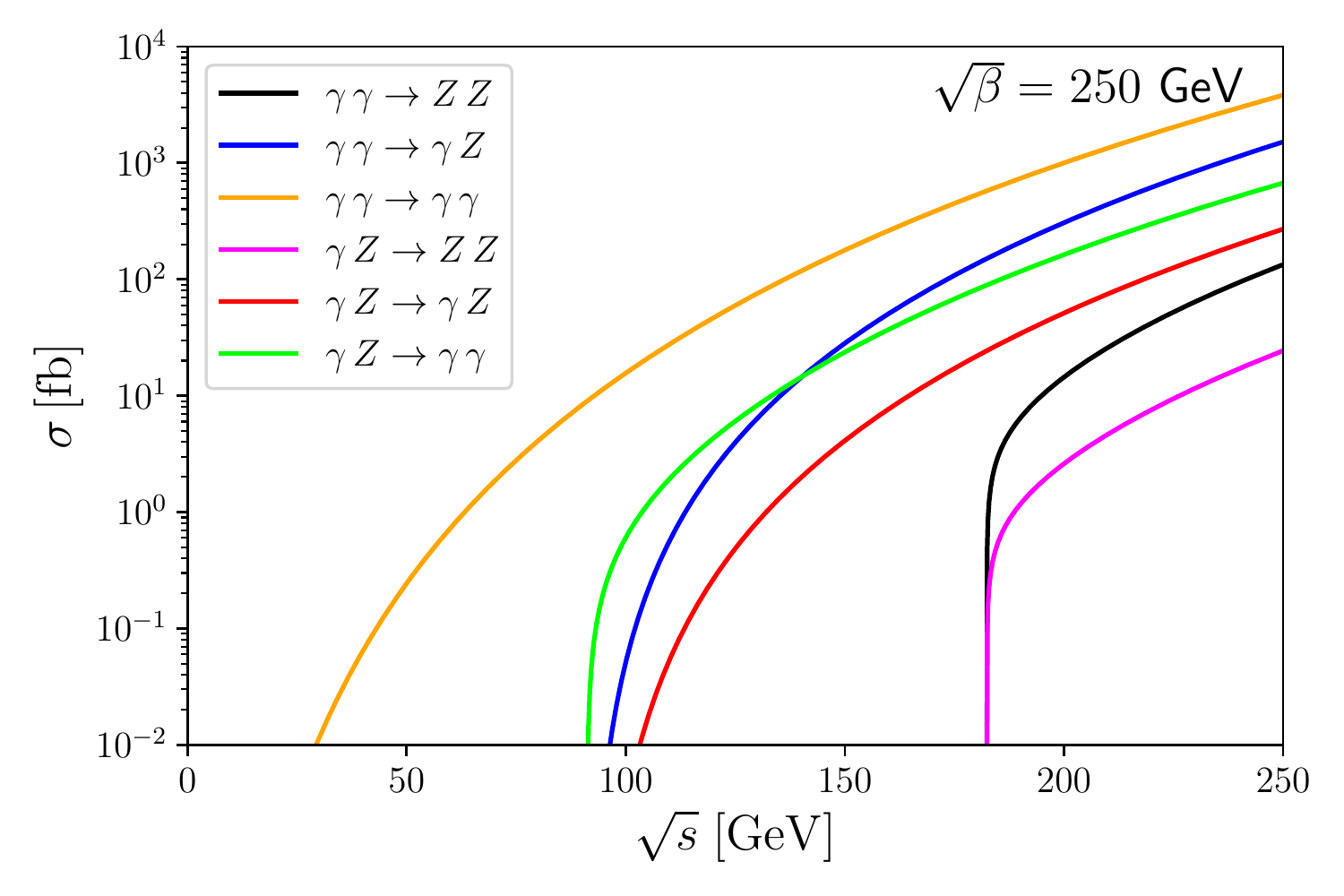}
\end{minipage} \hfill
\caption{Unpolarized total cross sections for selected processes (no angular cuts applied; evaluated in the c.m.); cf. Table~\ref{table:table_all}. Here, we set $\sqrt{\beta} = 250$~GeV for definiteness, but the scaling for other values can be easily performed via Eq.~\eqref{cs_all}.}
\label{fig_cs_all}
\end{figure}

Setting $x = m_Z^2/s$, the unpolarized differential cross section for the scattering $\gamma \gamma \rightarrow \gamma Z$ reads
\begin{align}\label{dcs_AAAZ}
\frac{d\sigma}{d\Omega} &= \frac{c_\theta^6 s_\theta^2 }{4096 \pi^2} \frac{s^3}{\beta^4} \left(1 - x \right)^3 \left[ \left(6 - 2 x^2\right) \cos^2\theta \right. \nonumber \\
& \left.  + \left(1 - x\right)^2 \cos^4\theta + 9 + 2 x + x^2\right] \, ,
\end{align}
which can be integrated to yield
\begin{eqnarray} \label{cs_AAAZ}
\sigma\left(\gamma \gamma \rightarrow \gamma Z\right)_Y & = & \frac{s^2_\theta c_\theta^6}{1920\pi} \left( \frac{s^3}{\beta^4} \right)  \left( 1- x \right)^3 \nonumber \\
& \times &  \left(21 + 3x + x^2\right) \, .
\end{eqnarray}
This result is shown in blue in Fig.~\ref{fig_cs_all} for $\sqrt{\beta} = 250$~GeV. If the non-linear hypercharge sector is indeed realized in nature, the expression above would provide the only tree-level contribution to the cross section, since this process cannot take place in the SM at this order. In fact, the first SM contribution is generated via fermion and W-boson loops with a cross section of $\sim 3 \times 10^{-4}$~fb shortly above threshold and peaking at $\sim 110$~fb at $\sim 750$~GeV~\cite{Dong}.

Another process of interest in a photon collider is $\gamma \, \gamma \rightarrow Z \, Z$, which, similar to $\gamma \, \gamma \rightarrow W^{+} \, W^{-}$, may be used to study the gauge structure of the SM as well as Higgs physics. As already mentioned, this process has no tree-level contribution in the SM -- the first non-trival amplitude arises through fermion and W-boson loops with a cross section of $\sim 20$~fb immediately after threshold and roughly saturating at $\sim 300$~fb for c.m. energies $\gtrsim 800$~GeV~\cite{tesla, Dicus, Jikia1}. In our non-linear extension the first non-zero contribution is at tree level, and the calculation of the (differential) cross section follows a rationale similar to  the one leading to Eq.~\eqref{cs_AAAZ}. The result is listed in Table~\ref{table:table_all} and shown in black in Fig.~\ref{fig_cs_all}.

Finally, let us briefly comment on $\gamma \, \gamma \rightarrow \gamma \, \gamma$, light-by-light (LbL) scattering. In Maxwell's linear electromagnetism this process is forbidden, but in the 1930s Heisenberg and Halpern~\cite{Halpern, Heisenberg} realized that quantum effects could induce it. In the 1950s a full calculation was presented~\cite{Karplus}, and the cross section was found to be $\sim 10^{-34}$~pb for visible light~\cite{Delbrueck, Tollis}. Only recently, it was proposed that this elusive process could be observed at the LHC in Pb-Pb collisions~\cite{Denterria} -- in fact, strong evidence for it has been reported by the ATLAS~\cite{Light1} and CMS collaborations~\cite{CMSlight}, being further confirmed by ATLAS~\cite{Light2}. The results are compatible with the SM prediction. As with the other photon-fusion processes previously discussed, LbL scattering takes place only at loop level in the SM. Our non-linear extension, on the other hand, allows for it to proceed already at tree level and with a potentially large cross section; cf. Fig.~\ref{fig_cs_all}.

The quartic vertices in Eq.~\eqref{lag_ZA} allow for a few more tree-level scattering processes involving exclusively neutral gauge bosons than we have explicitly mentioned above. For the sake of completeness, the differential cross sections for these processes are given in Appendix~\ref{app_dcs} and the respective total cross sections, without any angular cuts, can be written in a systematic way as
\begin{eqnarray} \label{cs_all}
\sigma & = & N \left( \frac{s^3}{\beta^4} \right) \kappa(x) F(x)\, 
\end{eqnarray}
with $x = m_Z^2/s$. Here, $N$ is a numerical factor, $\kappa(x)$ is a kinematic and phase-space factor, and further energy-dependent contributions are contained in $F(x)$. These results are summarized in Table~\ref{table:table_all} and shown in Fig.~\ref{fig_cs_all} for a reference value of $\sqrt{\beta} = 250$~GeV. The basic features are immediately salient: besides LbL scattering, all cross sections sharply rise after the respective thresholds and grow with increasing c.m. energy, as expected from the effective character of our hypercharge extension.

\setlength{\tabcolsep}{4pt}
\renewcommand{\arraystretch}{0.9}
\newcolumntype{C}[1]{>{\centering\arraybackslash}m{#1}}
\begin{table}[]
\centering
\begin{tabular}[t]{@{}|C{1.55cm}|C{0.65cm}|C{2.2cm}|C{3.cm}|C{-0.6cm}@{}}
\cline{1-4}
Process & $N$ & $\kappa(x)$ & $F(x)$   &   \\  [5pt]
\cline{1-4}
$\gamma \, \gamma \rightarrow \gamma \, \gamma$ & $\frac{7 c_\theta^8}{1280\pi}$ & 1 & 1    &   \\
[10pt]
\cline{1-4}
$\gamma \, \gamma \rightarrow \gamma \, Z$ & $\frac{s^2_\theta c_\theta^6}{1920\pi}$ & $\left( 1- x \right)^3$ & $21 + 3x + x^2$    &   \\ 
[10pt]
\cline{1-4}
$\gamma \, \gamma \rightarrow Z \, Z$ & $\frac{s^4_\theta c_\theta^4}{1280\pi}$ & $\sqrt{1 - 4x}$ & $7  - 26x + 27x^2$   &   \\ 
[10pt]
\cline{1-4}
$\gamma \, Z \rightarrow \gamma \, \gamma$ & $\frac{s^2_\theta c_\theta^6}{5760\pi}$ & $\left( 1- x \right)$ & $21 + 3x + x^2$  &   \\ 
[15pt]
\cline{1-4}
$\gamma \, Z \rightarrow \gamma \, Z$ & $\frac{s^4_\theta c_\theta^4}{2880\pi}$ & $\left( 1- x \right)^4$ & $21 + 6x + 16x^2 + 6x^3 + 6x^4$   &   \\ 
[15pt]
\cline{1-4}
$\gamma \, Z \rightarrow Z \, Z$ & $\frac{s^6_\theta c_\theta^2}{5760\pi}$ & $\left( 1- x \right) \sqrt{1 - 4x}$ & $21 - 75x + 98x^2 - 20x^3 + 6x^4$  &   \\ 
[10pt]
\cline{1-4}
\end{tabular}
\captionsetup{justification=centering}
\caption{Total cross sections for processes involving only quartic couplings of neutral gauge bosons; cf. Eq.~\eqref{cs_all} with $x = m_Z^2/s$. These results are shown in Fig.~\ref{fig_cs_all} for $\sqrt{\beta} = 250$~GeV.}
\label{table:table_all}
\end{table}

It is important at this point to contextualize our findings with recent results on anomalous quartic gauge couplings (aQGC). In fact, it is interesting to discuss experimental bounds and projected sensitivities on aQGC, as these may be translated into constraints on the parameter $\beta$ from different, but complementary, perspectives, ranging from past LHC runs to future lepton colliders. Particularly relevant is the discussion of recent results related to the anomalous vertices $\gamma \gamma \gamma Z$ and $\gamma \gamma \gamma \gamma$.

Anomalous quartic gauge couplings can be investigated with high precision at the LHC through $pp \rightarrow p X p$ processes, where X can be, for instance, $\gamma \gamma$ or $ \gamma Z$. In particular, it is interesting to focus on photon-induced processes in $pp$ collisions since these processes are very sensitive to aQGC and therefore new physics beyond the SM ({\it e.g.}, extended Higgs sectors or extra-dimensions).

For instance, the $\gamma \gamma \gamma Z$ interaction appears in the SM through fermion and W-boson loops. This anomalous vertex induces the rare decay $Z\rightarrow \gamma \gamma \gamma$, contributes to $e^+e^-\rightarrow \gamma \gamma \gamma $ and allows also for the $\gamma \gamma \rightarrow \gamma Z$ scattering. New physics appearing at a mass scale $\Lambda$, much heavier than the experimentally accessible energies $E$, can have its effects described via a low-energy effective field theory. The anomalous $\gamma \gamma \gamma Z$ interaction could then be parametrized by dimension-8 operators such as
\begin{align}\label{key}
	\mathcal{L}^{(1)}_{Z 3 \gamma} = \zeta \, F^{\mu \nu}F_{\mu \nu} F^{\rho \sigma} Z_{\rho \sigma} + \tilde{\zeta}  \, F^{\mu \nu}\tilde{F}_{\mu \nu} F^{\rho \sigma} \tilde{Z}_{\rho \sigma} \, .
\end{align}
This is an effective description, and we recover our model by making $\zeta = \tilde{\zeta} = -\frac{1}{8 \beta^2} s_\theta c_\theta^3$.


Baldenegro {\it et al.} studied in great detail the $\gamma Z$ production with intact protons in the forward region at the LHC using proton tagging~\cite{Baldenegro}. In this way, a sensitivity of $\zeta < 2 \times 10^{-13} \, \text{GeV}^{-4}$ could be established for the anomalous quartic gauge coupling $\gamma \gamma \gamma Z$ at an integrated luminosity of $300 \, \text{fb}^{-1}$. This improves the result obtained through the $Z \rightarrow \gamma \gamma \gamma$ measurement by about 3 orders of magnitude. This improvement in the anomalous coupling sensitivity would, in turn, translate into an improvement on the sensitivity of $\sqrt{\beta}$, putting it at the order of a few hundred GeV.

Recently, Inan and Kisselev studied very carefully the $\gamma \gamma \rightarrow \gamma Z$ scattering of photons produced by Compton backscattering at the CLIC and estimated the sensitivity to the anomalous quartic coupling $\gamma \gamma \gamma Z$~\cite{Kisselev1}. 
They used the following parametrization:
\begin{align}
	\mathcal{L}^{(2)}_{Z 3 \gamma} = g_1 \, F^{\rho \mu} F^{\alpha \nu} \partial_\rho F_{\mu \nu} Z_\alpha + g_2\, F^{\rho \mu} F^\nu_\mu \partial_\rho F_{\alpha \nu} Z^\alpha.
\end{align}
We can relate these coefficients with the previous ones through $g_1 = 8(\tilde{\zeta} - \zeta)$ and $ g_2 = 8 \tilde{\zeta}$. The authors considered both polarized and unpolarized $e^+e^-$ colisions at 1.5 and 3~TeV, obtaining exclusion limits on the aQGCs and comparing their results with the previous bounds obtained from $\gamma Z$ production at the LHC. The best bounds found by the authors for the couplings  $g_{1,2}$ were approximately $4.4 \times 10^{-14}\, \text{GeV}^{-4}$ and $5.1 \times 10^{-15}\, \text{GeV}^{-4}$ for the $e^+ e^-$ energies 1.5 and 3~TeV, respectively. They conclude that the sensitivities on the anomalous couplings obtained at CLIC are roughly 1 to 2 orders of magnitude stronger than the limits that can be obtained at the LHC. Such an improvement would be enough to put the sensitivity on our non-linear parameter $\sqrt{\beta}$ at the TeV scale. 

Let us now move on to the anomalous coupling $\gamma \gamma \gamma \gamma$. We can describe the non-linear effects on LbL scattering by means of the following effective Lagrangian:
\begin{align}\label{key}
	\mathcal{L}^{(1)}_{4 \gamma } = \zeta_1 F_{\mu \nu} F^{\mu \nu} F_{\rho \sigma} F^{\rho \sigma} + \zeta_2 F_{\mu \nu} F^{\nu \rho} F_{\rho \sigma} F^{\sigma \mu} \, .
\end{align}
This can be related to our description if we write it in a different, but equivalent basis given by:
\begin{align}\label{key}
	\mathcal{L}^{(2)}_{4 \gamma} = \xi F_{\mu \nu} F^{\mu \nu} F_{\rho \sigma} F^{\rho \sigma} + \tilde{\xi} F_{\mu \nu} \tilde{F}^{\mu \nu } F_{\rho \sigma} \tilde{F}^{\rho \sigma} \, ,
\end{align}
where the relation between the above parameters is given by $\xi = \zeta_1 + \frac{1}{2} \zeta_2$ and $\tilde{\xi}  = \frac{1}{4} \zeta_2$. We can recover our model if we take the particular combination $\xi = \tilde{\xi}= \frac{c_\theta^4}{32} \frac{1}{\beta^2}$.

Fichet {\it et al.} analyzed the sensitivities to the anomalous coupling $\gamma \gamma \gamma \gamma$ at the LHC through diphoton production with intact outgoing protons~\cite{Fichet1, Fichet2}. The reported limits at $14\, \text{TeV}$ with an integrated luminosity of $L = 300\,\text{fb}^{-1}$ on $\vert\zeta_1\vert$ and $\vert\zeta_2\vert$ were $1.5 \times 10^{-14} \, \text{GeV}^{-4}$ and $3.0 \times 10^{-14} \, \text{GeV}^{-4}$, respectively. For the High-Luminosity LHC (HL-LHC) the sensitivities obtained were a factor of 2 stronger. These results are strong and can put the sensitivity on our $\sqrt{\beta}$ at the TeV scale.

Inan and Kisselev examined the anomalous couplings $\gamma \gamma \gamma \gamma$ in the polarized LbL scattering at CLIC~\cite{Kisselev2}. Their results at $1.5 \, \text{TeV}$ were comparable with the bound obtained at HL-LHC stated above, but their results for $3 \, \text{TeV}$ were approximately 1 order of magnitude stronger, improving further the sensitivity on $\sqrt{\beta}$ but still keeping it at the TeV scale. A similar result could be found considering the best sensitivities on the anomalous couplings obtained through $\gamma \gamma \rightarrow Z Z$ in Ref.~\cite{Koksal2} and also through $Z \gamma \gamma$ production in Ref.~\cite{Gurkanli}, both considering $e^+ e^-$ collisions at 3 TeV at CLIC. 

Finally, let us conclude this section by reporting the latest experimental results on the anomalous couplings of interest here. A more general effective description including the nine independent dimension-8 operators respecting the $SU(2)_\text{L}\times U(1)_{\rm Y}$ gauge symmetry as well as charge conjugation and parity invariance can be found in Ref.~\cite{Eboli}. This effective description includes in particular
\begin{align}\label{key}
\mathcal{L} \supset F_{T,8} \, B_{\mu \nu} B^{\mu \nu} B_{\rho \sigma} B^{\rho \sigma} + F_{T,9} \, B_{\mu \nu} B^{\nu \rho} B_{\rho \sigma} B^{\sigma \mu} \, .
\end{align}
To the best of our knowledge, the strongest experimental bounds on these anomalous couplings are given by the very recent CMS results reported in Refs.~\cite{CMSanomalous1, CMSanomalous2, CMSanomalous3, CMSanomalous4}, considering different measurements in proton-proton collisions at $\sqrt{s} = 13\, \text{TeV}$ performed at the LHC. In particular, the strongest bounds on the anomalous couplings $F_{T,8}$ and $F_{T,9}$ are reported in Ref.~\cite{CMSanomalous4} and give $ \vert F_{T,8} \vert < 4.7 \times 10^{-13}  \, \text{GeV}^{-4}$ and $ \vert F_{T,9} \vert < 9.1 \times 10^{-13}  \, \text{GeV}^{-4}$. These coefficients are translated into our model by taking $F_{T,8} + F_{T,9}/2 = 1/32 \beta^2$ and $F_{T,9}/4 = 1/32 \beta^2$. Therefore, these experimental results put the bound on $\sqrt{\beta}$ at the order of a few hundred GeV.

Very recently, a work appeared on the arXiv~\cite{CMSphoton} in which the authors search for exclusive two-photon production via photon exchange in proton-proton collisions, $pp \rightarrow p \gamma \gamma p$, with intact protons using the CMS and TOTEM detectors at a center-of-mass energy of 13 TeV at the LHC. They report the following bounds on the anomalous four-photon coupling parameters: $\vert \zeta_1 \vert < 2.88 \times 10^{-13}\, \text{GeV}^{-4}$ and $\vert \zeta_2 \vert < 6.02 \times 10^{-13}\, \text{GeV}^{-4}$. This would give us a limit on $\sqrt{\beta}$ around the same order of magnitude as the result reported above.

The most recent and strong contribution to the subject was recently given by Ellis {\it et al.}~\cite{Ellisnovo}, constraining the non-linear scale of a BI extension of the SM to be $\gtrsim 5 \text{TeV}$ considering $gg \rightarrow \gamma \gamma $ at the LHC. The authors estimate the sensitivities at possible future pp colliders with $\sqrt{s} = 100\, \text{TeV}$ to be around $\gtrsim 20\, \text{TeV}$.
 
Therefore, we conclude that the LHC results can give very strong constraints on aQGC. These can be translated as bounds on $\sqrt{\beta}$ typically at the order of a few hundred GeV up to the TeV scale. Nevertheless, future colliders are expected to be able to supersede these constraints, consequently improving the sensitivity on $\sqrt{\beta}$.

	
\section{Conclusions} \label{sec_conclusion}
\indent

Motivated by recent results in the physics of electroweak monopoles, we investigated the consequences of a non-linear extension in the weak hypercharge sector in high-energy processes. The proposed extension is characterized by a parameter $\sqrt{\beta}$ with dimension of mass, which may be used to perform a Taylor expansion in $X = \frac{\mathcal{F}}{\beta^2} -\frac{\mathcal{G}^2}{2 \beta^4}$; cf. Eq.~\eqref{nonlinearlag}. After EW symmetry breaking, we obtain a series of quartic, dimension-8 effective operators involving the photon and Z-boson that are absent from the SM at tree level; cf. Eq.~\eqref{lag_ZA}.

In this context, we have analyzed a few interesting processes, namely, Z-decay and electron-positron annihilation, both resulting in three photons as final products, and Z-boson production via photon fusion. The first and most promising one, Z-decay, is a rare process occurring only at loop level in the SM, but induced at tree level by non-linear effects. The expected impact of the non-linear vertex on the branching ratio is a factor $\sim 3$ too small; cf. Eq.~\eqref{Gamma_Y}. This is due to the still-loose experimental upper bound on the branching ratio of Z-decay into three photons, which is 4 orders of magnitude larger than the value predicted by the SM.

Future $e^- e^+$ colliders, such as the ILC or FCC-ee, may operate at the Z-resonance and produce a large amount of Z-bosons -- up to a factor $10^5$ more than at LEP -- thereby dramatically increasing the statistics for measuring the products of Z-decay. We are therefore confident that the experimental upper limit on the branching ratio will be significantly improved in the near future, thus enabling us to set more stringent bounds on $\sqrt{\beta}$, readily excluding the range $\sqrt{\beta} \lesssim m_Z$; cf. Fig.~\ref{fig_beta_ZAAA}. We remark that, in a scenario where experiment reaches the level of the SM prediction, lower bounds $\sim 300$~GeV could be set.

The second process analyzed was electron-positron annihilation into three photons, also a relatively rare process. It is well described by QED, and the non-linear extension provides small corrections also at tree level; cf. Fig.~\ref{fig_eeAAA}. We have calculated the unpolarized cross sections of pure QED, pure non-linear and interference effects at the c.m. Since there is no tension between the predictions from QED and the experimental data, we have used the (relative) experimental uncertainties from LEP data for $e^- \, e^+ \rightarrow \gamma \gamma (\gamma)$ and $e^- \, e^+ \rightarrow 3 \, \gamma$ above the Z-pole to derive lower bounds on $\sqrt{\beta}$.

The process with two- and three-photon final states is well measured, but the non-linear effects, which contribute only to $e^- \, e^+ \rightarrow 3 \, \gamma$, are shadowed by the much larger QED contribution, $e^- \, e^+ \rightarrow 2 \, \gamma$. The data on exclusively three-photon final states is not so complete, but a conservative estimate delivers a somewhat better lower bound on $\sqrt{\beta}$. The non-linear effects are much smaller than the available precision and it was not possible to obtain viable bounds with the current experimental data, but we project that the necessary improvements may be within the reach of the next-generation lepton colliders.

Finally, we have also analyzed selected scattering processes involving exclusively neutral gauge bosons. The unpolarized tree-level cross sections may reach a few hundred fb at $\sqrt{s} = 200$~GeV for $\sqrt{\beta} = 250$~GeV; cf. Fig.~\ref{fig_cs_all}. These processes are good candidates to detect possible signatures from the non-linear extension in future experiments, given that they occur only at loop level in the SM, but are induced at tree level via Eq.~\eqref{lag_ZA}.

In this respect, we also reported recent results giving constraints on anomalous quartic gauge couplings obtained at the LHC considering neutral gauge-boson scatterings. We used them to estimate the corresponding limits on $\sqrt{\beta}$ and found that typically they give us bounds of a few hundred GeV. Furthermore, we analyzed the projections for these anomalous couplings in future lepton colliders and found that they improve the sensitivity on $\sqrt{\beta}$, putting it at the TeV scale.

Quite generally, we expect the non-linear effects to be heavily suppressed by $\sqrt{\beta}$ -- it could reach TeV energies depending on the underlying beyond-the-SM scenario. In this work, we have tried to constrain $\sqrt{\beta}$ with high-energy experiments, and we found that, in order to have any chance to detect such effects, very precise measurements are needed. A good example is the Z-decay in three photons, for which the (minute) SM contribution is generated at loop level, whereas the non-linear effects contribute already at tree level. This, together with the optimistic prospect of an improved upper limit on the branching ratio, makes this process a very promising way to search for the effects outlined in this work.

This can be contrasted to the situation of electron-positron annihilation: the non-linear effects are orders of magnitude smaller than the SM results and thus very hard to detect -- much like finding a needle in a haystack. We can see this by comparing the magnitudes of the cross sections in Eqs.~\eqref{cs_eeAAA_interf_207} and~\eqref{cs_eeAAA_Y207}, $\sim 0.9$~fb for $\sqrt{\beta} = 250$~GeV, with the size of the experimental uncertainties quoted in Ref.~\cite{L3_2002}, $\sim 0.5$~pb. Since the non-linear contributions are much smaller than the uncertainties involved, the only way to make them comparable [{\it \`a la} eq.~\eqref{eq_constraint_eeAA}] is by having $\sqrt{\beta} \lesssim \sqrt{s}$ to effectively enhance the non-linear effects.

Moreover, we remark that the experimental bounds on anomalous gauge couplings are being updated at a relatively fast pace. Their precise measurement is an extremely important task, since it provides a sensitive probe of new physics. We hope that future colliders will shine a new light on this issue, indicating the path to be followed on high-energy physics.


To conclude our contribution, we remark that a more general implementation of the non-linear extension of the electroweak sector is possible. Here, we have considered the $U(1)_{\rm Y}$ sector, but an analogous modification may be performed in the $SU(2)_{\rm L}$ sector. In this case, the already analyzed neutral sector would receive small modifications, but interesting non-linear effects would also be induced in the charged sector of the SM. This is the subject of another work to appear soon.

%
\begin{acknowledgments}
P.D.F. thanks especially G.P. de Brito for interesting discussions and technical support. He is grateful to Natanael C. Costa for discussions about Monte Carlo methods and W.B. de Lima and G. Pican\c{c}o for helpful discussions. P.C.M. is grateful to S.F. Amato, E. Polycarpo, and D. Kroff for their support. The authors are also grateful to the anonymous referee for his/her helpful suggestions. P.D.F. thanks the Brazilian scientific support agencies, CNPq and FAPERJ, for financial support. P.C.M. dedicates this paper to his daughter, Marina.
\end{acknowledgments}
%

\appendix

\section{Tree-level QED results for $e^- e^+ \rightarrow \gamma \gamma (\gamma)$} \label{app_A}
\indent

In Sec.~\ref{sec_eeAAA} we discussed the tree-level effects of the non-linear extension of the $U(1)_{\rm Y}$ sector in the process $e^- e^+ \rightarrow \gamma \gamma (\gamma)$. The experimental results from LEP included cross sections with final states of two and three photons subjected to detector cuts in energy and scattering angle, namely, $E_\gamma > 5$~GeV and $|\cos\theta_\gamma| < 0.96$~\cite{L3_1992, L3_2002}, so it is important to understand the tree-level expectation from QED to $e^- e^+ \rightarrow \gamma \gamma$ and $e^- e^+ \rightarrow \gamma \gamma \gamma$ under these conditions. 

We start with the simplest case, $e^- e^+ \rightarrow \gamma \gamma$. Since there are two identical particles in the final state and the reaction takes place at the c.m. , the two photons carry the same energy as the colliding electron. Assuming monochromatic beams with energies $\mathcal{O}(100 \, {\rm GeV})$, the outgoing photons automatically satisfy the energy cut. The tree-level differential cross section is given by the well-known result
\begin{equation} \label{dcs_qed_AA}
\frac{d\sigma^{2\gamma}_{\rm QED}}{d\cos\theta} = \frac{2\pi\alpha^2}{s} \left( \frac{1 + \cos^2\theta}{1 - \cos^2\theta} \right) \, .
\end{equation}
For two identical particles, the polar angle is confined to the range $0 \leq \cos\theta_\gamma \leq 1 - c_{\rm cut}$, and integrating Eq.~\eqref{dcs_qed_AA} in this range, we find\footnote{In order to keep track of the forward-backward enhancement in the ultra-relativistic limit it is usually imposed that $c_{\rm cut} = 2m_e^2/s$.}
\begin{equation} \label{cs_qed_AA}
\sigma^{2\gamma}_{\rm QED} = \frac{2\pi\alpha^2}{s} \left[ \log\left( \frac{2 - c_{\rm cut}}{c_{\rm cut}} \right) + c_{\rm cut} - 1  \right] \, .
\end{equation}
For LEP at $\sqrt{s} = 207$~GeV with $c_{\rm cut} = 0.04$, we get 9.6~pb. It is worth pointing out that the divergence in the forward-backward direction leads to a significant reduction of the total cross section even for small angular cuts.

Let us now move on to the more involved case of $e^- e^+ \rightarrow \gamma \gamma \gamma$. The typical amplitude is given in Eq.~\eqref{amp_eeAAA_0}, which must be added to other five similar contributions with permutations of the photon 4-momenta. If we define $p_{ij} = p_i \cdot q_j$ and $q_{ij} = q_i \cdot q_j$, the squared and spin-averaged amplitude can be written as
\begin{equation}
\langle \vert \mathcal{M}_{3\gamma} \vert^2 \rangle = \mathcal{Q} \left[ \, p_{11} \sum_{n=0}^{3} (p_1 \cdot p_2)^n Q_n + {\rm perm.}  \right]
\end{equation}
where ``perm.'' indicates that we must add the expression with the photon labels reshuffled. The pre-factor is
\begin{equation}
\mathcal{Q} = \frac{2 e^6}{ (p_{11}) (p_{12}) (p_{13}) (p_{21}) (p_{22}) (p_{23})} \, 
\end{equation}
and the terms in the sum are
\begin{subequations}
\begin{eqnarray}
Q_0 &=& p_{12} \Big[ p_{13} p_{21} p_{22}+p_{23} \big(p_{11} p_{22}+p_{23} (p_{22}-q_{12}) \nonumber \\
&+& p_{21} (p_{22}+q_{23})\big)\Big] - p_{11} p_{22} p_{23} q_{23}  \, , \\
Q_1 &=&   p_{12} \Big[ p_{13} p_{21}-p_{21} (p_{22}-4 p_{23})+p_{23} (p_{23}-q_{23})\Big] \nonumber \\
&+& p_{22} \Big[ p_{11} (-p_{22}+p_{23}+q_{23})  \nonumber \\
&-& p_{23} q_{12}+p_{22} q_{13}+p_{21} p_{23}\Big]  \, ,\\
Q_2 &=& - 2 p_{12} p_{21} - p_{22} (2 p_{21}-q_{12}+q_{13}+q_{23}) \, , \\
Q_3 &=&   p_{21} \, .
\end{eqnarray}
\end{subequations}

The final averaged squared amplitude can be symbolically recast in the form
\begin{equation}
\langle \vert \mathcal{M}_{3\gamma} \vert^2 \rangle = \frac{e^6}{E_{\rm cm}^2} \mathcal{C}(p_i, q_j) \, ,
\end{equation}
where we have expressed all dimensional parameters in terms of the c.m. energy -- in this way $\mathcal{C}(p_i, q_j)$ is effectively dimensionless. Taking into account the phase-space volume, cf. Eq.~\eqref{PI_3}, the integral to be solved is
\begin{equation} \label{I_qed}
\mathcal{I}_{\rm QED} = \int \mathcal{C}(p_i, q_j) \frac{d^3 {\bf q}_1}{E_1} \frac{d^3 {\bf q}_2}{E_2} \frac{d^3 {\bf q}_3}{E_3} \delta^4\left(\Sigma p_i - \Sigma q_j \right),
\end{equation}
but an analytical treatment is cumbersome, so we resort to numerical methods, which also facilitate the application of the detector cuts. The results of the Monte Carlo integral are listed in Table~\ref{table_integrals} for a few interesting values of the c.m. energy. The tree-level cross section for $e^- e^+ \rightarrow \gamma \gamma \gamma$ is ($e^2 = 4\pi \alpha$)
\begin{eqnarray} \label{cs_qed_AAA}
\sigma^{3\gamma}_{\rm QED} & = & \frac{\alpha^3}{48 \pi^2 s} \mathcal{I}_{\rm QED} \nonumber \\
& \simeq & 8 \times 10^{-3} \cdot \mathcal{I}_{\rm QED} \left(  \frac{200 \, {\rm GeV}}{\sqrt{s}} \right)^2 {\rm fb} \, .
\end{eqnarray}
Using $\sqrt{s} = 207$~GeV as an example, we have 0.285~pb.

\section{Interference and purely non-linear amplitudes for $e^- e^+ \rightarrow 3 \gamma$} \label{app_B}
\indent

Here we briefly present the results for the tree-level amplitudes discussed in Sec.~\ref{sec_eeAAA}. The interference amplitude between pure QED and the non-linear contributions may be written as
\begin{equation} \label{eq_app_interf}
\langle \vert \mathcal{M}_{\rm QED-Y} \vert^2 \rangle = \mathcal{H} \left[ \,\sum_{n=0}^{3} (p_1 \cdot p_2)^n H_n + {\rm perm.}  \right]
\end{equation}
with
\begin{eqnarray}
\mathcal{H} &=& \frac{c_\theta^2 e^4}{2 \beta^2 p_1 \cdot p_2 \left( p_{11} p_{12} p_{13} p_{21} p_{22} p_{23}\right)} \frac{\mathcal{H}_{\rm num}}{\mathcal{H}_{\rm den}}
\end{eqnarray}
and
\begin{subequations}
\begin{eqnarray}
\mathcal{H}_{\rm num} & = & 2 c_\theta^2 m_Z^2 \left(\Gamma_Z^2+m_Z^2\right) \nonumber \\
& - & \left(4 c_\theta^2+3\right) m_Z^2 (p_1 \cdot p_2) + 6 (p_1 \cdot p_2)^2 \, , \\
\mathcal{H}_{\rm den} & = & m_Z^4+\Gamma_Z^2 m_Z^2 \nonumber \\
& - & 4 m_Z^2 (p_1 \cdot p_2) + 4 (p_1 \cdot p_2)^2 \, .
\end{eqnarray}
\end{subequations}

The coefficients in Eq.~\eqref{eq_app_interf} are given by
\begin{subequations}
\begin{eqnarray}
H_0 &=& 	2 p_{22} \Bigg[ (p_{11})^2 \Big[ (p_{13})^3 p_{21} p_{22}  \nonumber \\
&+& p_{12} (p_{23})^2 (p_{21} q_{23}+p_{22} p_{23}) - p_{12} p_{13} p_{23} \Big(p_{12} p_{21}  \nonumber \\
&+& p_{21} (q_{23}-2 p_{23})+p_{22} (p_{23}-q_{13})\Big) \Big] \nonumber \\
& - & p_{11} p_{12} p_{13} p_{21} p_{23} (p_{21} q_{23}+p_{22} p_{23}) \nonumber \\
&-& (p_{12})^2 (p_{13})^2 (p_{21})^2 p_{23} \Bigg] \, , \\
H_1 &=& -p_{22} \Bigg\{ 2 p_{12} p_{13} (p_{21})^2 p_{22} q_{13} \nonumber \\
& + & (p_{11})^2 p_{23} \Big[2 (p_{12})^2 (p_{23}+q_{13}) \nonumber \\
&+& p_{12} \Big(p_{13} (p_{21}+p_{22})-2 p_{21} q_{23}\Big)+p_{13} p_{21} p_{22}\Big] \nonumber \\
&+& p_{11} p_{21} \Big[2 (p_{13})^2 p_{21} p_{22} + p_{12} p_{13} \Big(2 (p_{23} q_{12}+q_{13} q_{23}) \nonumber \\
& + & p_{21} (p_{23}-2 q_{23})\Big)+2 p_{12} p_{23} q_{13} q_{23}\Big] \Bigg\} \, , \\
H_2 &=& p_{11} p_{21} \Big[p_{13} p_{22} (2 p_{11} q_{23} +p_{23} q_{12} ) \nonumber \\
&+& p_{12} p_{23} (p_{13} q_{12}+2 p_{21} q_{23})\Big] \, .
\end{eqnarray}	
\end{subequations}

Finally, the purely non-linear amplitude is given by
\begin{equation} \label{eq_app_Y}
\langle \vert \mathcal{M}_{\rm Y} \vert^2 \rangle = \mathcal{J} \left[ \, \sum_{n=0}^{3} (p_1 \cdot p_2)^n J_n + {\rm perm.}  \right]
\end{equation}
where 
\begin{equation} 
\mathcal{J} = \frac{c_\theta^4 e^2}{2\beta^4 (p_1 \cdot p_2)^2} \frac{\mathcal{J}_{\rm num}}{\mathcal{J}_{\rm den}} \, ,
\end{equation}
with
\begin{subequations}
\begin{eqnarray}
\mathcal{J}_{\rm num} & = & 2 c_\theta^4 m_Z^2 \left(\Gamma_Z^2+m_Z^2\right) \nonumber \\
& - & 6 c_\theta^2 m_Z^2 (p_1 \cdot p_2) + 5 (p_1 \cdot p_2)^2  \, , \\
\mathcal{J}_{\rm den} & = & m_Z^4+\Gamma_Z^2 m_Z^2 \nonumber \\
& - & 4 m_Z^2 (p_1 \cdot p_2) + 4 (p_1 \cdot p_2)^2 \, .
\end{eqnarray}	
\end{subequations}

The coefficients in Eq.~\eqref{eq_app_Y} are given by
\begin{subequations}
\begin{eqnarray}
J_0 &=&  3 (p_{11})^2 p_{22} p_{23} q_{23}-6 p_{11} p_{12} p_{22} p_{23} q_{13} \nonumber \\
&+& 3 p_{12} p_{13} (p_{21})^2 q_{23}  \, , \\
J_1 &=&  3 p_{11} q_{23} (p_{21} q_{23}-2 p_{22} q_{13}) \nonumber \\
&+& (p_{11})^2 (q_{23})^2+(p_{21})^2 (q_{23})^2  \, , \\
J_2 &=&   q_{12} q_{13} q_{23} \, .
\end{eqnarray}	
\end{subequations}

\vspace{0.1cm}

\section{Differential cross sections for neutral gauge boson scatterings} \label{app_dcs}
\indent 

In this Appendix, we report the unpolarized differential cross-sections for the scattering of neutral gauge bosons 
in the non-linear extension considered in this work. In the following, we are using $\beta_Z \equiv \sqrt{1 - 4 m_Z^2 / E_{\rm cm}^2 }$.

\begin{itemize}
	
\item $\gamma \gamma \rightarrow \gamma \gamma$
\begin{align}\label{key}
	\frac{d\sigma}{d\Omega} = \frac{E_{\rm cm}^6 c_\theta^8 \left(3+\cos^2\theta\right)^2}{ 4096 \pi^2 \beta^4}.
\end{align}		
	
	
\item $\gamma \gamma \rightarrow \gamma Z$ 
\begin{align}\label{key}
	\frac{d\sigma}{d\Omega} &= \frac{c_\theta^6 s_\theta^2 \left(E_{\rm cm}^2 - m_Z^2\right)^3 }{4096 \pi^2 \beta^4 E_{\rm cm}^4} \left[ \left(6 E_{\rm cm}^4 - 2 m_Z^4\right) \cos^2\theta \right. \nonumber \\
	& \left. + 9 E_{\rm cm}^4 + \left(E_{\rm cm}^2 - m_Z^2\right)^2 \cos^4\theta + 2 E_{\rm cm}^2 m_Z^2 + m_Z^4\right].
\end{align}


\item  $\gamma \gamma \rightarrow Z Z$
\begin{align}\label{key}
	\frac{d\sigma}{d\Omega} &= \frac{c_\theta^4 s_\theta^4 E_{\rm cm}^2  \beta_Z}{4096 \pi ^2 \beta ^4}	\Big[  9 E_{\rm cm}^4 + E_{\rm cm}^4 \beta_Z^4 \cos^4\theta \nonumber \\
	&+ 6 E_{\rm cm}^4  \beta_Z^2 \cos^2\theta  -32 E_{\rm cm}^2 m_Z^2+40 m_Z^4 \Big].
\end{align}	


\item  $\gamma Z \rightarrow \gamma \gamma$
\begin{align}\label{key}
	\frac{d\sigma}{d\Omega} &= \frac{c_\theta^6 s_\theta^2 \left(E_{\rm cm}^2 - m_Z^2\right)}{6144 \pi^2 \beta^4}  \left[\left(6 E_{\rm cm}^4 - 2 m_Z^4\right) \cos^2\theta \right. \nonumber \\
	& \left. + 9 E_{\rm cm}^4 + \left(E_{\rm cm}^2 - m_Z^2\right)^2 \cos^4\theta + 2 E_{\rm cm}^2 m_Z^2 + m_Z^4\right].	
\end{align}


\item  $\gamma Z \rightarrow \gamma Z$
\begin{align}\label{key}
	\frac{d\sigma}{d\Omega} &= \frac{	c_\theta^4 s_\theta^4  \left(E_{\rm cm}^2 - m_Z^2\right)^4}{49152 \pi ^2 \beta^4 E_{\rm cm}^{10}} \Big[ 99 E_{\rm cm}^8+20 E_{\rm cm}^6 m_Z^2 \nonumber \\
	&+74 E_{\rm cm}^4 m_Z^4+20 E_{\rm cm}^2 m_Z^6+\left(E_{\rm cm}^2-m_Z^2\right)^4 \cos4 \theta  \nonumber \\
	&-8 m_Z^4 \left(E_{\rm cm}^2-m_Z^2\right)^2 \cos3 \theta  \nonumber \\
	&-8 m_Z^4 \left(11 E_{\rm cm}^4+2 E_{\rm cm}^2 m_Z^2+7 m_Z^4\right) \cos\theta \nonumber \\
	&+4 \big(7 E_{\rm cm}^8-4 E_{\rm cm}^6 m_Z^2+4 E_{\rm cm}^4 m_Z^4 \nonumber \\
	&-4 E_{\rm cm}^2 m_Z^6+7 m_Z^8\big) \cos2 \theta  +35 m_Z^8 \Big].
\end{align}


\item  $\gamma Z \rightarrow Z Z$
\begin{align}\label{key}
	\frac{d\sigma}{d\Omega} &= 	\frac{c_\theta^2 s_\theta^6 \beta_Z \left(E_{\rm cm}^2 - m_Z^2\right)}{6144 \pi^2 \beta^4 E_{\rm cm}^4} \left[9 E_{\rm cm}^8 - 30 E_{\rm cm}^6 m_Z^2 \right. \nonumber \\
	& \left. + 45 E_{\rm cm}^4 m_Z^4 + \left(E_{\rm cm}^4 - 5 E_{\rm cm}^2 m_Z^2 + 4 m_Z^4\right)^2 \cos^4\theta \right. \nonumber \\
	& \left. + 2 E_{\rm cm}^4 \beta_Z^2  \left(3 E_{\rm cm}^4 + m_Z^4\right) \cos^2\theta \right].
\end{align}


\end{itemize}

\end{document}